\documentclass[journal,onecolumn,draftclsnofoot,11pt]{IEEEtran}
\usepackage{amsthm}
\usepackage{amsmath,graphicx}
\usepackage{amssymb}
\usepackage{subfig}
\usepackage{graphicx}
\usepackage{graphics}
\usepackage{epstopdf}
\usepackage{amsmath}
\usepackage{multirow}
\usepackage{bm}
\usepackage{enumerate}
\usepackage{mathdots}
\usepackage{todonotes}
\usepackage{algorithmicx}
\usepackage{algpseudocode}
\usepackage{algorithm}
\usepackage{subfig}
\usepackage{amsthm}
\usepackage{acronym}
\usepackage{amsfonts}
\usepackage{times}
\usepackage{booktabs}
\usepackage[flushleft]{threeparttable}
\usepackage{latexsym}
\usepackage{dsfont}
\usepackage{cite}
\usepackage{verbatim}
\usepackage{epstopdf}
\usepackage{color}

\def\bb0{{\mathbb{0}}}

\def\ba{{\mathbf{a}}}
\def\bb{{\mathbf{b}}}

\def\bm{{\mathbf{m}}}

\def\b0{{\mathbf{0}}}

\def\bA{{\mathbf{A}}}
\def\bB{{\mathbf{B}}}

\def\bH{{\mathbf{H}}}

\def\bR{{\mathbf{R}}}

\def\bbE{{\mathbb{E}}}

\def\cA{\mathcal{A}}

\def\cN{\mathcal{N}}

\def\sf0{{\mathsf{0}}}

\newcommand{\sref}[1]{{Section}~\ref{#1}}

\definecolor{purple(x11)}{rgb}{0.63, 0.36, 0.94}
\definecolor{cadmiumgreen}{rgb}{0.0, 0.42, 0.24}

\begin{document}

\title{Hybrid MIMO Architectures for Millimeter Wave Communications: Phase Shifters or Switches?}
\author{Roi M\'{e}ndez-Rial, Cristian Rusu, Nuria Gonz\'{a}lez-Prelcic, Ahmed Alkhateeb, and \\Robert W. Heath Jr.
\thanks{R. M\'{e}ndez-Rial, C. Rusu and Nuria Gonz\'{a}lez-Prelcic are with the University of Vigo, Spain, (email: \{roi,crusu,nuria\}@gts.uvigo.es).
A. Alkhateeb and R. W. Heath Jr. are with The University of Texas at Austin, Austin, TX, USA (email: \{aalkhateeb,rheath@utexas.edu)\}. 
N. Gonz\'alez-Prelcic would like to acknowledge support from the Spanish Government and the European Regional Development Fund (ERDF) under projects TACTICA and COMPASS (TEC2013-47020-C2-1-R).
R. Heath would like to acknowledge support from the National Science Foundation under Grant No. NSF-CCF-1319556 and a gift from Nokia.}}

\maketitle

\begin{abstract}
Hybrid analog/digital MIMO architectures were recently proposed as an alternative for fully-digital-precoding in millimeter wave (mmWave) wireless communication systems. This is motivated by the possible reduction in the number of RF chains and analog-to-digital converters. In these architectures, the analog processing network is usually based on variable phase shifters. In this paper, we propose hybrid architectures based on switching networks to reduce the complexity and the power consumption of the structures based on phase shifters. We define a power consumption model and use it to evaluate the energy efficiency of both structures. To estimate the complete MIMO channel, we propose an open loop compressive channel estimation technique which is independent of the hardware used in the analog processing stage.  We  analyze the performance of the new estimation algorithm for hybrid architectures based on phase shifters and switches. Using the estimated, we develop two algorithms for the design of the hybrid combiner based on switches and analyze the achieved spectral efficiency. Finally, we study the trade-offs between power consumption, hardware complexity, and spectral efficiency for hybrid architectures based on phase shifting networks and switching networks.  
Numerical results show that architectures based on switches obtain equal or better channel estimation performance to that obtained using phase shifters, while reducing hardware complexity and power consumption.
For equal power consumption, all the hybrid architectures provide similar spectral efficiencies.
\end{abstract}

\section{Introduction} \label{sec:intro}
Communication over millimeter wave (mmWave) frequencies will be a key feature of the fifth generation (5G) cellular networks \cite{pi2011introduction,Rapp5G,Boccardi2014,Rangan2014,Andrews2014}. Thanks to the large bandwidth channels potentially available, mmWave communication can meet the high peak data rate requirements of next-generation wireless systems. In the USA, the Federal Communication Commission has just recognized the potential of mmWave technologies for mobile cellular in a proposed rulemaking \cite{FCC15-138}. Another potential advantage of mmWave communication is its low latency \cite{Rapp5G}, which is essential for many 5G applications, like wearable networks \cite{Venugopal2015,Pyattaev2015}, autonomous robots, and connected or self-driving cars \cite{Lu2014}. MmWave wireless communication has also been considered for many other applications including local and personal area networks \cite{11ad,WirelessHDSrandard2007}, joint vehicular communication and radar \cite{Heddebaut2010,Kumari2015}, and simultaneous energy/data transfer \cite{Bi2015,Khan2015}.  

One of the key architectural features of mmWave is the use of large antenna arrays at both the transmitter and the receiver \cite{Rapp5G,mmWavePrecoding_Samsung,Pi2011}. These arrays are used  to provide array gain and obtain enough link margin for wide area operation. Unlike lower frequency MIMO systems, the large arrays combined with high cost and power consumption of the mixed analog/digital signal components makes it difficult to assign an RF chain per antenna, and perform all the signal processing in the baseband \cite{Singh2009,ElAyach2014,mmWaveTutorial2015}. This motivates the development and analysis of new transceiver structures and their impact on MIMO signal processing including precoding, combining, and channel estimation \cite{Alkhateeb2014d}. 

Analog beamforming is an approach that relies entirely on RF domain processing to reduce the number of RF chains\cite{Wang2009,Chen2011,Hur2013}. The beamforming is implemented using networks of analog phase shifters that change the relative phases of the signals feeding the antennas to steer the transmit/receive beams in the desired directions \cite{Wang2009,Chen2011,Hur2013}. Ultimately, the performance of analog beamforming is limited by quantization of the phase angles and the support of only single stream MIMO transmission.

\begin{figure*}[t]
	\centering
	\includegraphics[width=0.8\linewidth]{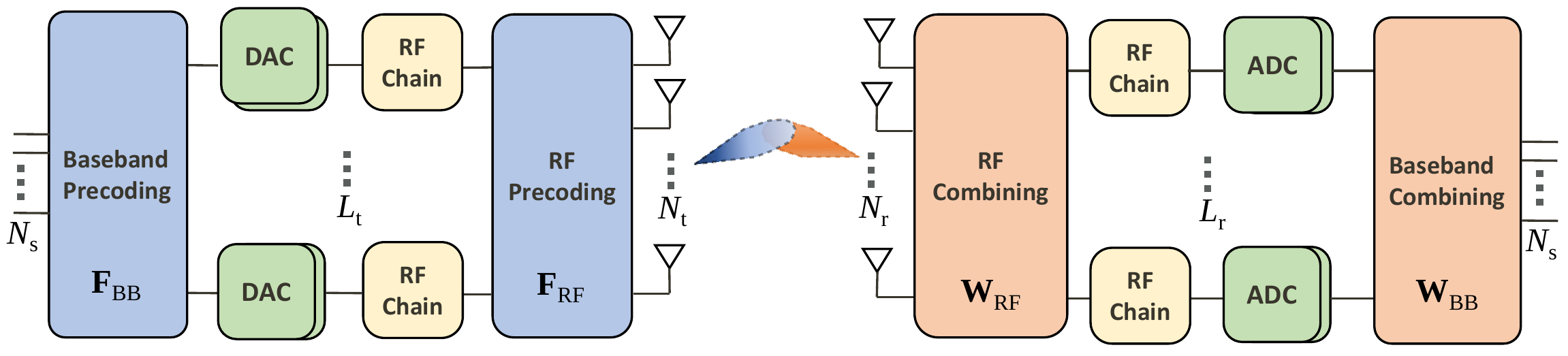}%
	\caption{Hybrid MIMO architecture for mmWave communication. Precoding and combining are divided between analog and digital domains. The number of RF chains at the transmitters/receivers are therefore much less than the number of antennas.}
	\label{fig:hybrid}
\end{figure*}

An alternative approach to reduce the number of RF chains is hybrid analog/digital architectures \cite{Molisch2005,Venkateswaran2010,Ayach2014,Alkhateeb2013,Alkhateeb2014, Han2015,ElAyach2013}. In hybrid architectures, MIMO processing is split between the analog and digital domains to reduce the number of required transceivers, as shown in Fig.~\ref{fig:hybrid}. Thanks to the extra digital processing, more degrees of freedom are available for the design of the hybrid precoders compared with analog-only beamformers, allowing them to support  multi-stream and multi-user transmission. 

Analog and hybrid MIMO architectures proposed so far for mmWave communications are usually based on phased arrays, though there is some work on alternative lens based architectures \cite{brady:taps12}. The analog network of phase shifters feeding the antennas, however, is not a simple circuit at mmWave \cite{Poon2012}. Practical phased arrays use finite precision phase shifters, which may lack accuracy, and make it difficult to finely steer the beams and place nulls. Increasing the number of bits in the phased array  leads to a higher complexity and higher power consumption. Although power consumption can be reduced using passive phase shifters instead of active ones, the insertion losses are higher \cite{Poon2012}. This is not desirable, given that the combining network after the phase shifters also introduces insertion losses. Further, while only one low-noise amplifier (LNA) per antenna is typically needed in conventional MIMO receivers, mmWave receivers based on phased arrays require larger numbers of LNAs to achieve the same signal-to-noise ratio (SNR) at the input of the RF chains. The effect of these practical limitations motivates the research for alternatives to phase shifters in mmWave hybrid architectures.


In this paper, we propose \textbf{new hybrid architectures based on  networks of switches} instead of phase shifters in the analog processing stage. The objective of the proposed architectures is to reduce cost, complexity, and power consumption of mmWave MIMO systems, while incurring small loss in the system performance. We focus on the design of the combiners at the receiver, though many concepts also extend to transmit precoding. We consider two types of  switch-based hybrid architectures: (i) an architecture where only one antenna is selected for each RF chain, and (ii) an architecture where a subset of antennas is selected, and their outputs are combined, for each RF chain. While the first architecture with an antenna per RF chain has lower complexity, the second one makes better use of the offered array gain. This motivates characterizing the trade-off between the achievable rate and the power consumption of these proposed architectures.

To compare the different switch-based and phase shifter-based hybrid architectures, we develop \textbf{a power consumption model for mmWave MIMO receivers}. The developed model provides a general framework for calculating the consumed power in the hybrid architectures as a function of the numbers of antennas and RF chains. Using this model and approximated power consumption numbers from recent RF circuit designs \cite{Kraemer2009, Fonte2011, Liu2011, Chang2012, SundeepCTW13, Hong2015, Sung2015,Dortz2014,Dong2014,SLee2014,Miyahara2014,Janssen2013,Kull2013,Tabasy2013,Shettigar2012,Poon2012,Franc2013,Natarajan2011,Li2013,Uzunkol2012,Kuo2012,Kim2012,Yu2010,Kim2010,Liu2009,Belov2012,Schmid2014,Ziegler2000}, we show that switches based architectures can yield reasonable reduction in the power consumption, especially with large antenna arrays.

To exploit the new proposed architectures, we develop new mmWave channel estimation algorithms to configure the hybrid precoders and combiners.  Channel estimation is difficult  in mmWave because of the large channel dimensions and the low receive SNR before beamforming design \cite{mmWaveTutorial2015}. With hybrid architectures, the mmWave channel estimation problem becomes more difficult and architecture-dependent because the channel is seen at the baseband through the RF lens. Exploiting channel sparsity, though,  channel estimation at mmWave can be formulated as a compressive channel estimation problem \cite{Bajwa2010}, where different compressed sensing (CS) tools can be leveraged to estimate the channels \cite{Ramasamy2012,Alkhateeb2014,Lee2014, Berraki2014, Alkhateeb2015}.

Compressive estimation of sparse spatial channels at mmWave was first proposed in \cite{Ramasamy2012} for an analog beamforming architecture. The algorithm estimates a predetermined number of paths using a coarse grid for the possible spatial frequencies, then refines those estimates using Newton's method. In \cite{Alkhateeb2014} a hybrid architecture is considered, and adaptive compressed sensing tools from \cite{Malloy2014} were leveraged to iteratively estimate the dominant paths of mmWave channels. In \cite{Lee2014, Berraki2014}, compressed sensing with random training sequences were adopted also for channel estimation in hybrid architectures. An extension to multi-user mmWave systems was provided in \cite{Alkhateeb2015}, where multiple users estimate their channels simultaneously. The solutions in \cite{Alkhateeb2014,Lee2014, Berraki2014, Alkhateeb2015} all focus on  mmWave channel estimation and the training signals design assuming phase shifters are used for the analog beamforming stage. They cannot be used for switches. 

We propose \textbf{an open loop strategy for downlink mmWave channel estimation}. The proposed algorithm is independent of the hardware used in the analog processing stage and thus can be applied to either phase shifter or switching networks. It also incorporates hybrid constraints at both the transmitter and receiver during training, and leverages mmWave channel sparsity in the angular domain. For each of the proposed architectures, we design efficient training beamformers and combiners that lead to appropriate CS measurement matrices, and derive bounds on their performance. For the noiseless case, we find the optimal number of training measurements that minimizes the Welch bound on the coherence of the equivalent measurement matrix  \cite{Welch1974}. Using the normalized mean squared error (NMSE) as a performance metric, we show by simulations that the developed training sequence designs and channel estimation algorithms with  switch-based architectures can achieve comparable performance to that obtained with phase shifter based architectures, while requiring lower power consumption.

Given an estimate of the channel, it is possible to optimally configure the hybrid precoders and combiners. In \cite{Molisch2005},  hybrid precoders were designed in general MIMO systems for multiplexing gain maximization; interference management was further incorporated in \cite{Venkateswaran2010}. Leveraging mmWave channel sparsity, \cite{Ayach2014} developed an orthogonal matching pursuit based solution for hybrid precoders in mmWave systems, assuming perfect channel knowledge. Designs that do not rely on matching pursuit have been proposed in \cite{Sohrabi2015,Rusu2015}. The solutions in \cite{Ayach2014,Sohrabi2015,Rusu2015}, though, considered hybrid architectures with analog phase shifters. For the switch-based hybrid architectures, the design of the analog precoders/combiners is related to the antenna (subset) selection problem, which is a classical topic in the MIMO literature (see e.g. \cite{SanNos2004,MolWin2004} and the references therein). Prior work on antenna subset selection, however, focused on fully-digital and not hybrid switch-based architectures. Therefore, new precoding/combining algorithms need to be developed for the proposed switch-based hybrid architectures.

In this paper, we  devise \textbf{adaptive strategies for the design of the switch-based hybrid architectures with antenna subset selection}. The proposed solutions incorporate hybrid constraints into the greedy algorithms in \cite{Gharavi-alkhansari2004}. Equipped with a channel estimator and a means of deriving the precoding and combining matrices, we analyze the performance of the different architectures in numerical simulations. We compare the achieved spectral efficiencies of the proposed switch-based hybrid precoding/combining algorithms with that obtained using fully-digital architectures and  phase-shifters hybrid analog/digital constraints. From these results, we conclude that architectures based on switches can achieve similar spectral efficiencies for equal power consumption. Further, when comparing all architectures operating with the same number of RF chains, we find that those architectures based on switches require lower power consumption at the cost of a small loss in the array gain, which slightly impacts the obtained spectral efficiency.

In summary, we provide a complete design of several hybrid mmWave architectures based on switching networks for the analog processing stage. In \sref{sec:sysmodel}, we define the system model and describe four different MIMO architectures that make use of antenna switches and subarrays.  Then we develop a power consumption model in \sref{sec:powermodel}, and use it to craft a comparison between the power efficiency of the different proposed architectures. We present the design of a novel compressive channel estimator in Section \ref{sec:train}, which can be used with either phase shifters or switches in the analog processing. Based on the channel estimates, we propose algorithms for the design of the analog and digital combining matrices in Section \ref{sec:combine}. Then in \sref{sec:results} we compare the different architectures in terms of channel estimation error and spectral efficiency, showing when the switch-based architecture is preferred. We make some concluding remarks in \sref{sec:conclusion}. 

We use the following notation throughout this paper: bold lowercase $\ba$ is used to denote column vectors, bold uppercase $\bA$ is used to denote matrices, non-bold letters $a,A$ are used to denote scalar values, and caligraphic letters $\cA$ to denote sets. Using this notation, $|a|$ is the magnitude of a scalar, 
$\|\ba\|_2$ is the $\ell_2$ norm, $\|\ba\|_{0}$ is the $\ell_0$ pseudo-norm,  $\|\bA\|_F$ is the Frobenius norm, 
$\sigma_k(\bA)$ denotes the $k^{\text{th}}$ singular value of $\bA$ in decreasing order, $\mathrm{tr}(\bA)$ denotes the trace, 
$\bA^*$ is the conjugate transpose, $\bA^T$ is the matrix transpose,  $\bA^{-1}$ denotes the inverse of a square matrix, 
$\ba_k$ is the $k^{\text{th}}$ entry of $\ba$, $|\cA|$ is the cardinality of set $\cA$. 
$\bA \otimes \bB$  is the Kronecker product of $\bA$ and $\bB$. We use the notation $\cN(\bm,\bR)$ to denote a complex circularly symmetric Gaussian random vector with mean $\bm$ and covariance $\bR$. We use $\bbE$ to denote expectation.

\section{System model} \label{sec:sysmodel}
Consider the single-user mmWave hybrid MIMO system in Fig.~\ref{fig:hybrid}. In this paper we focus on the downlink and specifically consider antenna selection at the mobile station (MS) where the complexity and power consumption limitations are especially important. While we describe the general model including hybrid operation at both the BS and the MS, we focus on the combining operation at the MS since the base station typically has less restrictive power constraints. 

The transmitting BS is equipped with $N_\text{t}$ antennas and $L_\text{t}$ RF chains, while the receiving MS has $N_\text{r}$ antennas and $L_\text{r}$ RF chains.
$N_\text{s}$ data streams are transmitted from the BS to the MS assuming $N_\text{s} \leq L_\text{t} \leq N_\text{t}$ and $N_\text{s} \leq L_\text{r} \leq N_\text{r}$. 
The transmitter applies a hybrid precoder $\mathbf{F}$ to the symbol vector $\mathbf{s} \in \mathbb{C}^{N_\text{s} \times 1}$ with $\mathbb{E}[\mathbf{ss}^*] = \frac{1}{N_\text{s}}\mathbf{I}$.
The hybrid precoder $\mathbf{F}=\mathbf{F}_\text{RF} \mathbf{F}_\text{BB}$ is composed of an RF precoder  $\mathbf{F}_\text{RF} \in \mathbb{C}^{N_\text{t} \times L_\text{t}}$, and a baseband digital precoder $\mathbf{F}_\text{BB} \in \mathbb{C}^{L_\text{t} \times N_\text{s}}$.
The discrete-time transmitted signal is given by $\mathbf{x}=\mathbf{Fs}$.

We consider a narrowband frequency-flat channel model represented by the channel matrix $\mathbf{H} \in \mathbb{C}^{N_\text{r} \times N_\text{t}}$, with $\mathbb{E} \left [ \| \mathbf{H}\|_F^2 \right ]=N_\text{t}N_\text{r}$.
Assuming perfect synchronization, the received signal can be written as 
\begin{equation} 
\mathbf{r}=\sqrt{\rho}\mathbf{HFs}+\mathbf{n} ,
\end{equation}
where $\rho$ represents the average received power  and $\mathbf{n} \in \mathbb{C}^{N_\text{r}\times 1}$ is the noise vector with $\mathcal{CN}(0,\sigma_n^2)$ entries. 
The MS applies a hybrid combiner $\mathbf{W}$ to the received signal so that the processed received signal is
\begin{equation}
\mathbf{y}=\sqrt{\rho}\mathbf{W}^*\mathbf{H}\mathbf{F}\mathbf{s}+\mathbf{W}^*\mathbf{n} .
\end{equation}

The hybrid combiner $\mathbf{W}=\mathbf{W}_\text{RF} \mathbf{W}_\text{BB}$ is composed of an RF combiner $\mathbf{W}_\text{RF} \in \mathbb{C}^{N_\text{r} \times L_\text{r}}$ and a baseband combiner $\mathbf{W}_\text{BB} \in \mathbb{C}^{L_\text{r} \times N_\text{s}}$.
The RF precoder and combiner are implemented in  analog, so that the precoding and combining matrices $\mathbf{F}_\text{RF}$ and $\mathbf{W}_\text{RF}$ are subject to specific constraints depending on the hardware used to implement the analog precoder/combiner.

\begin{figure}[htbp]
\centering
\subfloat[A1: Architecture with variable phase shifters]{\includegraphics[width=0.38\linewidth]{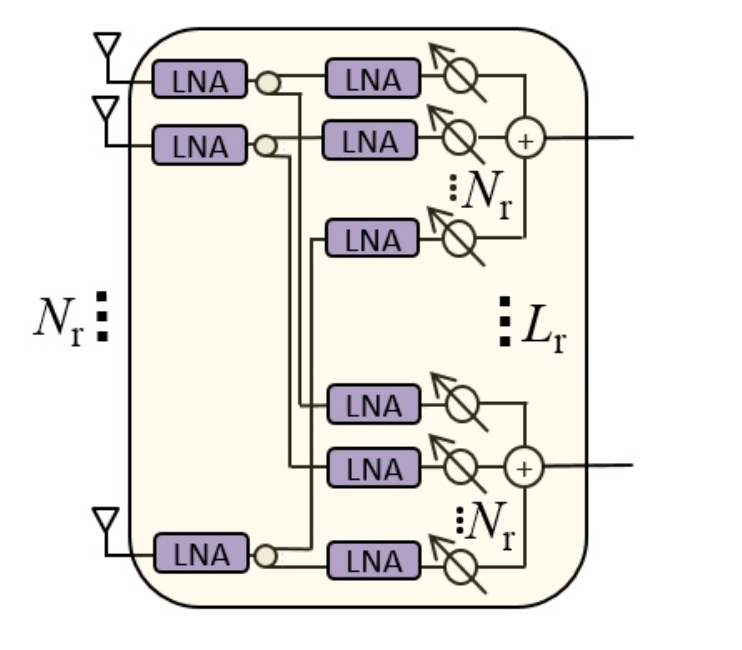}%
\label{fig:A1}}
\hfil
\subfloat[A2: Architecture with variable phase shifters in subsets of antennas]{\includegraphics[width=0.28\linewidth]{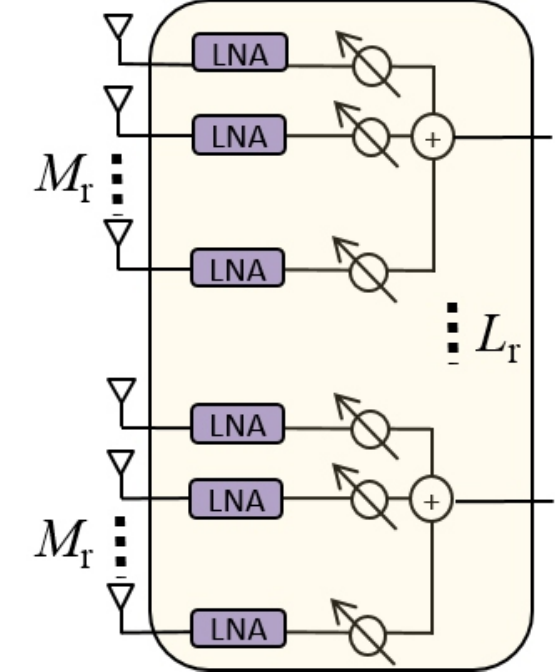}%
\label{fig:A2}}\\
\subfloat[A3: Antenna selection with analog combining]{\includegraphics[width=0.35\linewidth]{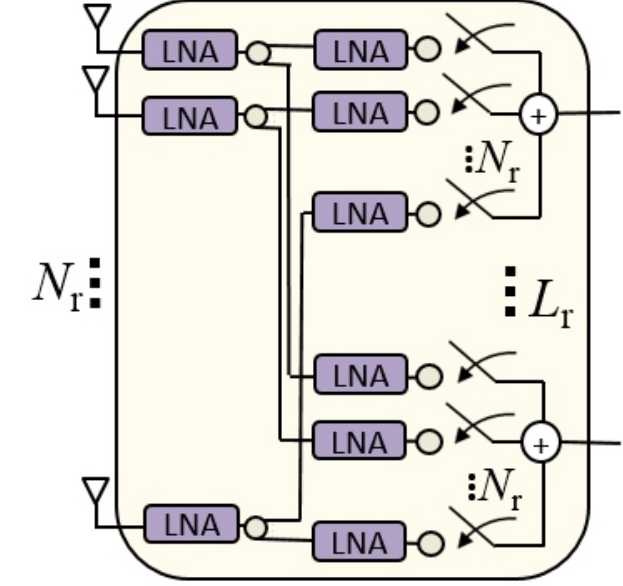}%
\label{fig:A3}} 
\hfil
\subfloat[A4: Antenna selection in subsets with analog combining ]{\includegraphics[width=0.28\linewidth]{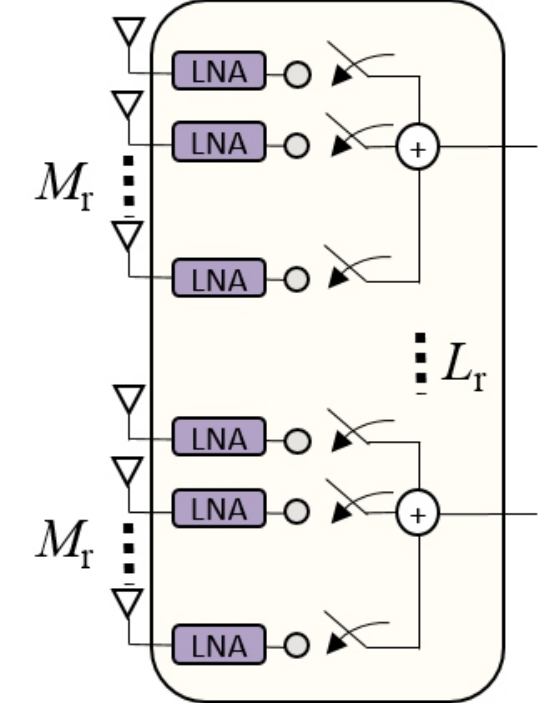}%
\label{fig:A4}}\\
\subfloat[A5: Antenna selection]{\includegraphics[width=0.35\linewidth]{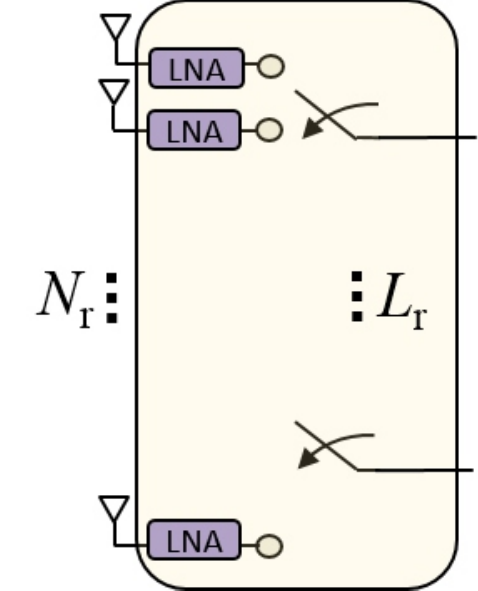}%
\label{fig:A5}} 
\hfil
\subfloat[A6: Antenna selection in subsets]{\includegraphics[width=0.24\linewidth]{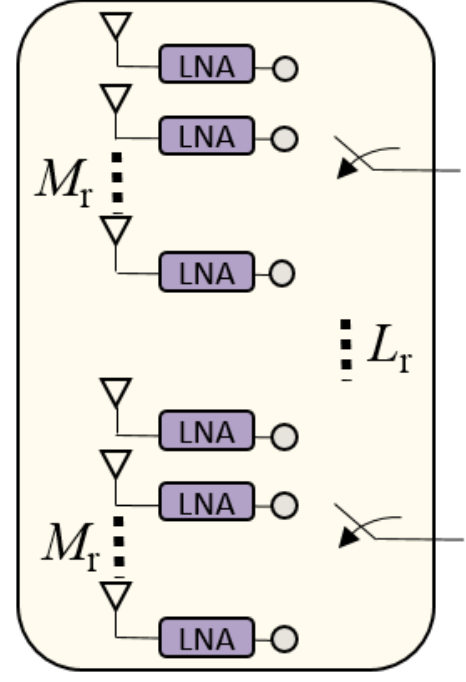}%
\label{fig:A6}} \\
\caption{Analog architectures for the RF combiner.}
\label{fig_sim}
\end{figure}

We consider two hybrid architectures that use phase shifters based on previous work. 
Further, we propose four new hybrid precoding  architectures based on switches. Fig. \ref{fig_sim} shows the block diagrams corresponding to all these architectures, which are further explained in the following.

\paragraph*{A1) Phase shifting network} Each transceiver is connected to each antenna through a network of phase shifters.
Assuming infinite resolution phase shifters, the incoming signals are combined before feeding the RF chain.
Each transceiver achieves full array gain emulating an antenna with high aperture.
An RF precoder/combiner implemented with a network of variable phase shifters imposes the constraint of unit norm entries in $\mathbf{W}_\text{RF}$ and $\mathbf{F}_\text{RF}$.
The set of feasible combining vectors, e.g. columns of $\mathbf{W}_\text{RF}$, is given by $\mathcal{A}_1 = \{ \mathbf{x} \in \mathbb{C}^{N_\text{r} \times 1} \; | \; |\mathbf{x}_i|=1 \}$.

\paragraph*{A2) Phase shifting network in subsets} Each transceiver is routed to a subset of $M_\text{r}=N_\text{r}/L_\text{r}$ antennas through an analog preprocessing network of analog phase shifters. This is essentially an array-of-subarrays hybrid architectures, which assume that each RF is connected to a unique subset of the antennas \cite{Han2015,ElAyach2013}. Compared to architecture A1, the maximum gain of each transceiver is reduced by a factor $1/L_\text{r}$.
The complexity of this architecture, however,  is lower and it requires only $N_\text{r}$ RF paths and analog phase shifters instead of the $N_\text{r} \times L_\text{r}$ needed in A1.
This solution is interesting for the BS in multiuser scenarios, since the antenna grouping reduces the mutual coupling and interference between users \cite{Han2015}.
The set of feasible RF combining vectors is given by
\begin{equation}
\mathcal{A}_2 = 
\left \{  \begin{bmatrix}
\hat{\mathbf{x}}\\ 
0\\ 
\vdots\\ 
0
\end{bmatrix}
\cup 
\begin{bmatrix}
0\\ 
\hat{\mathbf{x}}\\ 
\vdots\\ 
0
\end{bmatrix}
\cup 
\hdots
\begin{bmatrix}
0\\ 
0\\ 
\vdots\\ 
\hat{\mathbf{x}}
\end{bmatrix}
\in \mathbb{C}^{N_\text{r} \times 1} 
\left | 
\begin{array}{l}
\hat{\mathbf{x}}\in\mathbb{C}^{N_\text{r}/L_\text{r}\times 1}\\ 
|\hat{\mathbf{x}}_i|=1
\end{array}\right.
\right \}.
\end{equation}

\paragraph*{A3) Switching network with analog combining}
In this new architecture, each transceiver is routed to each antenna through an network of switches.
A3 is equivalent to A1, replacing each variable phase shifter by a simple switch.
Each transceiver selects a subset of $N_\text{A3}$ antennas, between 1 and $N_\text{r}$, whose received signal will be non coherently combined before feeding the RF chain, which could result in a degradation of the SNR.
Therefore, designing an optimal combiner presents challenging issues.
The set of feasible RF combiners is given by $\mathcal{A}_3 = \{\mathbf{x}\in\mathcal{B}^{N_\text{r}\times 1}\}$, where $\mathcal{B}$ is the binary set $\{0,1\}$.

\paragraph*{A4) Switching network in subsets with analog combining}
In this case  each antenna is routed to a single switch, avoiding the need of splitters and the following LNAs, which results in a lower power consumption than the previous architecture. Only $N_\text{A4}$ of the antennas in every subset of size $M_\text{r}$ are  combined, with $1\leq N_\text{A4} \leq M_\text{r}$.
The set of feasible combiners is given by 
\begin{equation}
\mathcal{A}_4 = 
\left \{  \begin{bmatrix}
\hat{\mathbf{x}}\\ 
0\\ 
\vdots\\ 
0
\end{bmatrix}
\cup 
\begin{bmatrix}
0\\ 
\hat{\mathbf{x}}\\ 
\vdots\\ 
0
\end{bmatrix}
\cup 
\hdots
\begin{bmatrix}
0\\ 
0\\ 
\vdots\\ 
\hat{\mathbf{x}}
\end{bmatrix}
\in \mathcal{B}^{N_\text{r} \times 1}
\left | 
\hat{\mathbf{x}}\in\mathcal{B}^{N_\text{r}/L_\text{r}\times 1}
\right.
\right \}.
\end{equation}

\paragraph*{A5) Switching network}
Antenna selection in a switching network is a low-cost low-complexity alternative to capture many of the advantages of MIMO systems \cite{Sanayei2004}. 
Each transceiver can be connected to any antenna through a $N_\text{r}$:1 antenna switch. 
Only $L_\text{r}$ antennas are active at a time. To avoid large insertion losses, this architecture is more suited for small arrays to be used in mobile devices for example. In this case a
switch can connect to any antenna in the array, but there will be only 2-4 antennas. 
The effective gain of the system is limited by the number of active antennas.
The RF combining/precoding matrices become selection matrices routing $L_\text{r},L_\text{t}$ antennas to the corresponding RF chain.
The set of feasible RF combiners is given by 
$\mathcal{A}_5 = \{ \mathbf{x} \in \mathcal{B}^{N_\text{r} \times 1} \; | \; \|\mathbf{x}\|_0 =1 \}$ with $\mathcal{B}= \{ 0,1 \}$.
Each column of $\mathbf{W}_\text{RF},\mathbf{F}_\text{RF}$ is a binary vector with a single one and zeros elsewhere.
Since only one switch can be connected to a single antenna simultaneously, the number of nonzero elements in each row is $\leq 1$.
The switch introduces insertion losses into the signal, but the insertion losses due to the combining network used with the phase shifters are avoided. 

\paragraph*{A6) Switching network in subsets}
Only one antenna per subset is selected so that the insertion losses due to the combiner are also avoided. 
It is an interesting alternative to A5 if the array size is large, since it could be infeasible to 
switch between a large number of antennas.
Considering predefined subsets of contiguous antennas, the set of feasible combiners is given by
\begin{equation}
\mathcal{A}_6 = 
\left \{  \begin{bmatrix}
\hat{\mathbf{x}}\\ 
0\\ 
\vdots\\ 
0
\end{bmatrix}
\cup 
\begin{bmatrix}
0\\ 
\hat{\mathbf{x}}\\ 
\vdots\\ 
0
\end{bmatrix}
\cup 
\hdots
\begin{bmatrix}
0\\ 
0\\ 
\vdots\\ 
\hat{\mathbf{x}}
\end{bmatrix}
\in \mathcal{B}^{N_\text{r} \times 1}
\left | 
\begin{array}{l}
\hat{\mathbf{x}}\in\mathcal{B}^{N_\text{r}/L_\text{r}\times 1}\\ 
\| \hat{\mathbf{x}}_i \|_0=1
\end{array}\right.
\right \}.
\end{equation}

The goal of this work is to analyze the trade-offs between each architecture in terms of power consumption, channel estimation capability and spectral efficiency performance. Since the power consumption considerations are much more pressing at the MS than at the BS we analyze the power consumption at the receiver side, and also focus on channel estimation and hybrid combining algorithms.

\section{Power consumption model} \label{sec:powermodel}
Understanding power consumption is important for characterizing tradeoffs between the different hybrid architectures. 
In this section, we develop an approximate power consumption model for the architectures in Fig.~\ref{fig_sim}.
Our aim is to compare the architectures proposed in the previous section in terms of power consumption for different values of the array size and the number of RF chains. 
Since exact computation of the dissipated power is difficult in general, we approximate the power consumption of each hardware component and give reasonable arguments for the choice we make.
We focus the analysis only on the receiver side.
A similar model can be derived for the transmitter replacing the low noise amplifiers (LNA) by power amplifiers (PA), and the ADCs with DACs.

We denote as $P_{\text{LNA}}$ the power consumed by a single LNA and $P_{\text{ADC}}$ the power consumed by a single ADC. 
We assume that all the architectures use the same kind of ADCs and LNAs. 
Recent work on the development of low power LNAs for the $60$ GHz band in CMOS technology \cite{Kraemer2009, Fonte2011, Liu2011,Chang2012}, show that power consumption for an LNA at $60$ GHz is in the range of $4.6$ to $86$ mW. 
Other results in the literature, for example the mmWave picocellular system in \cite{SundeepCTW13}, estimated the LNA consumption at 20 mW which is within our range. 
In this paper we will assume $P_\text{LNA} = 20$ mW.

Regarding power consumption of mmWave ADCs, we focus on devices that can be used in a $60$ GHz system with at least a $500$ MHz  bandwidth and  an effective resolution (ENOB) larger than $4$ bits.
Table~\ref{ADCconsumption} shows ADC power consumption versus maximum sampling frequency and effective number of bits (ENOB) for different designs proposed between 2012 and 2015. 
The values show the power consumption of prototype ADCs that can be found in the recent scientific literature. 
Although they give a clear trend in the development of low power ADCs, to be conservative about the power consumption of commercially available mmWave converters in the future years, we consider a reference value of $P_\text{ADC} = 200$ mW, that includes the conversion of the I and Q components (note that some of the integrated designs referenced in the table already include two converters). The choice of the reference value is difficult in this case since the values in Table \ref{ADCconsumption} present high variability and therefore a threshold cannot be clearly discerned. We choose a conservative (high) value since the designs in Table \ref{ADCconsumption} are not commercial products and as such we might expect these values to be relatively optimistic (low), as compared to the power consumption of the final working devices. Furthermore, previous work \cite{SundeepCTW13} in the same context of mmWave communications also considers a similar reference value.

\begin{table}[htbp]
\begin{center}
\begin{tabular}{|l|c|c|c|c|}
\hline
& & Maximum  &   &  Power \\ 
Reference& Year & sampling  & $\text{ENOB}_{hf}$ & consumption \\
& & rate & &   \\
& & [Gs/s]  & [bits] & [mW]\\ \hline
Hong et al.\cite{Hong2015} & 2015 & 1.7 & 8.21 & 15.4 \\ \hline
Sung et al.\cite{Sung2015} & 2015 & 1.6 & 9 & 17.3 \\ \hline
Le Dortz et al.\cite{Dortz2014} & 2014 & 1.62 & 7.68 &  93\\ \hline
Dong et al.\cite{Dong2014} & 2014 & 3.20 & 11.77 & 240 \\ \hline
Lee et al.\cite{SLee2014} & 2014 & 1.0 & 8.2 & 19.8 \\ \hline
Miyahara et al.\cite{Miyahara2014} & 2014 & 2.2 & 5.92 & 27 \\ \hline
Janssen et al.\cite{Janssen2013} & 2013 & 3.6 & 8.01 & 795 \\ \hline
Kull et al.\cite{Kull2013} & 2013 & 1.2 & 6.24 & 3.1 \\ \hline
Tabasy et al.\cite{Tabasy2013} & 2013 & 3 & 4.56 & 79.1 \\ \hline
Shettigar et al.\cite{Shettigar2012} & 2012 & 3.6 & 11.49 & 15 \\ \hline
\end{tabular}
\caption{ADC power consumption versus sampling frequency and effective number of bits.}
\label{ADCconsumption}
\end{center}
\end{table}


RF phase shifters are key components in integrated mmWave phased arrays.
Some important considerations regarding the phase shifter design are the noise figure, power consumption, insertion loss, loss variation over the range of shift, linearity, resolution and bandwidth.
Phase shifters at mmWave frequencies can be classified into: (i) active phase shifters, such as reflective, loaded line, switched delay; and (ii) passive phase phase shifters, such as cartesian vector modulator, LO-path phase shifter, and phase-oversampling vector modulator \cite{Poon2012}\cite{Franc2013}.
Active phase shifters have a small footprint but generally cause nonlinearity problems resulting in a high noise figure.
Passive phase shifters occupy a larger area and incur in large insertion losses.
Although there is not a significant noise trade-off between active and passive phase shifters \cite{Natarajan2011}, the power consumption with passive phase shifters is lower, which make them preferable for applications requiring mobility and autonomy.
The main challenge associated with passive phase shifters is the dependency of the insertion losses on the phase shift setting.
To compensate this effect, variable gain amplifiers (VGAs) with integrated phase-compensated technique are usually added following the passive phase shifter \cite{Natarajan2011}.

\begin{table}[htbp]  
  \centering  
	\begin{threeparttable}
    \begin{tabular}{|l|c|c|r|r|r|}
		\hline
    Reference   &  Year & Freq  & Resolution & Peak Gain  & Power  \\
		&   & [Ghz]  & [$^\circ$]-[bits] & [dB]  & [mW]  \\
		\hline
    Li et al.\cite{Li2013} & 2013 & 58-65 & 11.25 - 5& -5.4  & 31.2 \tnote{a} \\
    Uzunkol et al.\cite{Uzunkol2012} & 2012 & 60-67 & 45 - 3& 6.5 & 15 \\
    Kuo et al.\cite{Kuo2012} & 2012 & 57-62 & 22.5 - 4& $28$ & 45 \\
    Kim et al.\cite{Kim2012} & 2012 & 55-78 & 22.5 - 4& 14 & 108 \\
    Natarajan et al.\cite{Natarajan2011} & 2011 & 57-65 & 11.25 - 5 & 58  & 57\\
    Yu et al.\cite{Yu2010} & 2010 & 58-64 & 22.5 - 4& $14$ & 52\\
    Kim et al.\cite{Kim2010} & 2010 & 53-65 & 11.25 - 5& 12.5 & 50 \\		
		\hline				
    \end{tabular}%
		\begin{tablenotes}
            \item[a] LNA not included.            
    \end{tablenotes}
		\end{threeparttable}
	\caption{Power consumption versus phase resolution for mmWave variable phase shifters proposed in the recent lieterature.  }			
	\label{PSconsumption}%
\end{table}%

The power consumed by the phase shifters depends on the type and on the resolution of the quantized phases. 
Table~\ref{PSconsumption} summarizes recent work on variable phase shifters at mmWave frequencies in the literature.
Note that the power consumption of the LNA together with the phase shifter is included except for \cite{Li2013}.
Power consumptions for one element of the receiver phased array is considered in \cite{Yu2010,Natarajan2011,Kuo2012}.
Based on Table \ref{PSconsumption}, we assume a reference power consumption value $P_{\text{PS}} = 30$ mW. 
We focus on phase shifters with at least $4$ bits of angular resolution.

The RF combining network in the array can be active or passive. 
Power dissipation and linearity constraints favor full passive-combining approaches \cite{Natarajan2011}.
However, including active combining stages can improve array efficiency reducing the passive combining loss.
In a phase array, noise from the RF front end starts to dominate the output noise as more elements of the array are turn on.

MmWave switches have to be capable of switching at nanosecond or sub-nanosecond speeds. 
Different implementations have been proposed recently for mmWave applications \cite{Liu2009,Belov2012} which exhibit low insertion loss ($\approx 1$dB) and good isolation properties. 
The power consumption of the switch is low \cite{Schmid2014,Ziegler2000}, we assume a value of $P_\text{SW} = 5$ mW.

The RF chain block in all the architectures includes the following devices: a mixer, a local oscillator, a low pass filter and a base band amplifier. The power consumed by the RF chain can be written as
\begin{equation}
\label{eq:power_save}
P_{\text{RFC}}=P_{\text{mixer}}+P_{\text{LO}}+P_{\text{LPF} }+P_{\text{BB amp}}.
\end{equation}
Some reference data for the power consumption of other components of the RF chain are given in \cite{SundeepCTW13}: mixer 19 mW, LO buffer 5 mW, filter 14 mW, baseband amplifier 5 mW.
From this data, we consider an approximate reference power consumption of $40$ mW for the complete RF chain. 

In \cite{BBprocessor}, a baseband precoder/combiner is designed and implemented in CMOS, with a power consumption of $243$ mW.
This power consumption obviously depends on the particular precoder/combiner design included in the transmitter/receiver. 
To be able to obtain a comparison of the power dissipated for all the architectures we will also assume that the baseband processor consumes the same power as a single ADC, that is, $P_{\text{BB}}=P_{\text{ADC}}$.

Based on the block diagrams, the power dissipated by the hybrid combining architectures can be written respectively as
\begin{align}
& P_\text{A1}=N_\text{r}(L_\text{r}+1)P_{\text{LNA}}+N_\text{r}L_\text{r}P_\text{PS}+L_\text{r}(P_{\text{RFC}}+P_{\text{ADC}}) \\
& \qquad +P_{\text{BB}}
\label{power with phase shifters} \\
& P_\text{A2}=N_\text{r}P_{\text{LNA}}+N_\text{r}P_\text{PS}+L_\text{r}(P_{\text{RFC}}+P_{\text{ADC}})+P_{\text{BB}},
\label{power with phase shifters in subsets}\\
& P_\text{A3}=(N_\text{r}+L_\text{r}N_\text{A3})P_\text{LNA}+L_\text{r}N_\text{A3}P_\text{SW}\\
& \qquad + L_\text{r}(P_\text{RFC}+P_\text{ADC})+P_\text{BB},
\label{power with switches and combiner}\\
& P_\text{A4}=L_\text{r}N_\text{A4}(P_\text{LNA}+P_\text{SW})+L_\text{r}(P_{\text{RFC}}+P_{\text{ADC}})+P_{\text{BB}},
\label{power with switches and combiner in subsets}\\
& P_\text{A5}=L_\text{r}(P_\text{LNA}+P_\text{SW})+L_\text{r}(P_{\text{RFC}}+P_{\text{ADC}})+P_{\text{BB}},
\label{power with switches}\\
& P_{\text{A6}}=L_\text{r}(P_\text{LNA}+P_\text{SW})+L_\text{r}(P_{\text{RFC}}+P_{\text{ADC}})+P_{\text{BB}},
\label{power with switches and subsets}
\end{align}
Note that the power consumption for the architecture (A2) does not depend on the array size, but on the number of RF chains. 
The power consumption of a complete antenna array with one RF chain per antenna is
\begin{equation}
P_\text{D}=N_\text{r}(P_{\text{LNA}}+P_{\text{RFC}}+P_{\text{ADC}})+P_{\text{BB}}.
\end{equation}
The percentage of power consumption reduction that can be achieved with the hybrid architectures respect to that of the complete system can be written as
\begin{equation}
\eta_i = P_i/P_\text{D}, \;\;\; i=1\hdots 4 .
\end{equation}

To compare power consumption we write these equations in terms of a single power value $P_{\text{ref}} = 20$ mW. 
In this way:
\begin{align*}
P_{\text{LNA}} & =P_{\text{ref}} \\
P_{\text{ADC}} & =10P_{\text{ref}} \\
P_{\text{RF chain}} & =2 P_{\text{ref}} \\
P_{\text{BB}} & =10P_{\text{ref}} \\
P_{\text{PS}} & =1.5P_{\text{ref}} \\
P_{\text{SW}} & =0.25P_{\text{ref}} \\
\end{align*}

\begin{figure}[!htb]
\begin{minipage}[b]{1.0\linewidth}
  \centering
  \centerline{\includegraphics[width=0.5\linewidth]{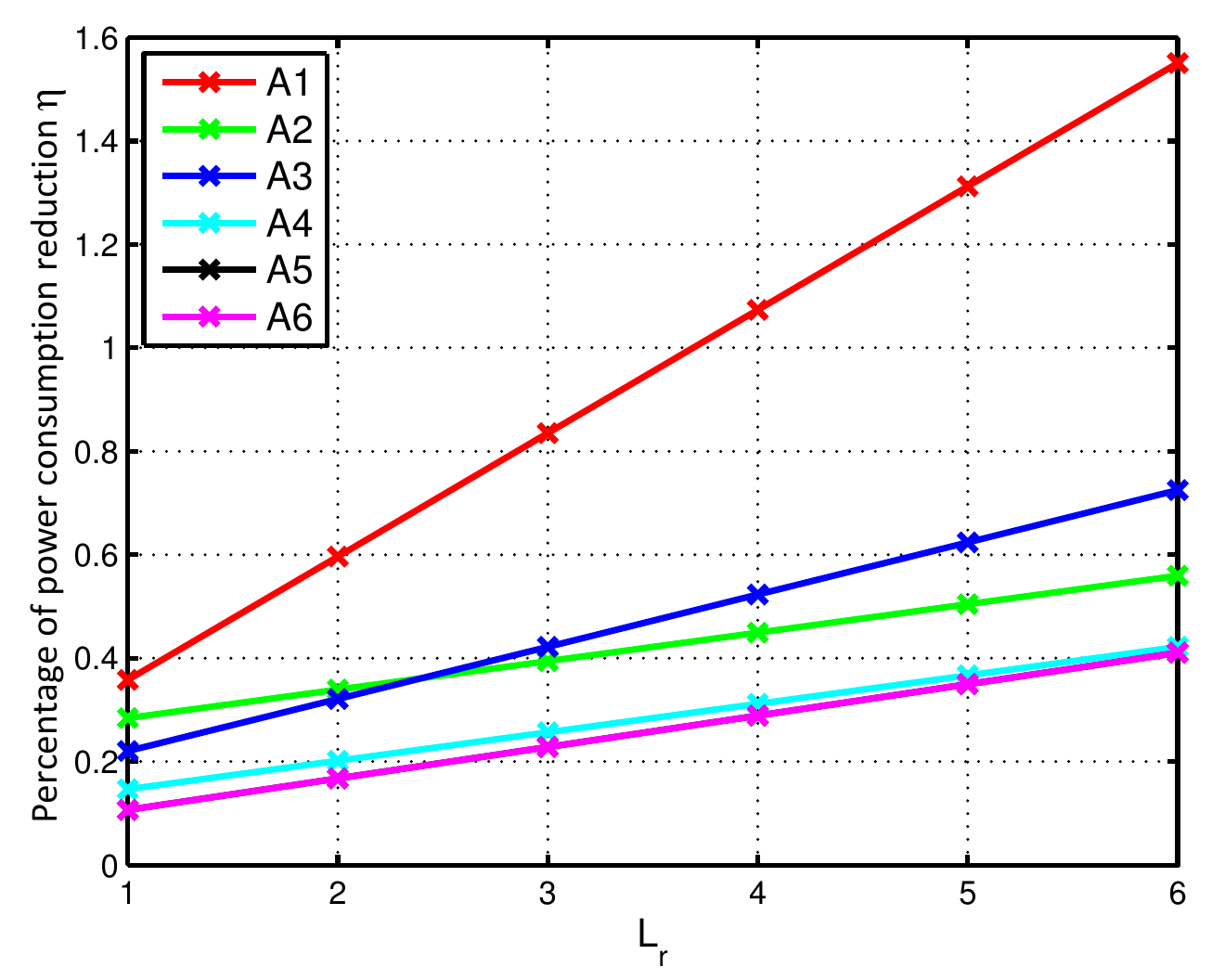}}
  \caption{Power consumption reduction of the RF front-end of the hybrid architectures as a function of the number of RF chains respect to a complete digital system with one RF chain per antenna. $N_\text{r}=16$ receive antennas are considered. The reference power consumption of a complete array with one transceiver is $P_\text{D}= 4360$mW.}
\label{powerReduction}
\end{minipage}
\end{figure}

Fig.~\ref{powerReduction} shows the power consumption reduction $\eta$ as a function of the number of RF chains for $N_\text{r}=16$.
For the architectures A3 and A4, we consider that half of the switches in each subset are active, $P_\text{A3}=0.5N_\text{r}$ and $P_\text{A4}=0.5N_\text{r}/L_\text{r}$.


\section{Downlink compressed sensing based channel estimation} \label{sec:train}
Acquiring channel knowledge at the receiver is useful for designing the hybrid beamformers and combiners. 
In this section, we describe the sparse multipath channel model and present a compressed sensing based channel estimation approach that works with all the architectures.
Although we focus on single-user mmWave systems, the proposed approach can also work for multi-user (MU) MIMO systems.
The main advantage of this technique in MU-MIMO is that all the MSs can simultaneously estimate their downlink channels.
The consequence is a reduction of the training overhead compared to the user-specific approaches \cite{Alkhateeb2014,Wang2009,Hur2013} where the complexity scales with the number of MSs.

\subsection{Sparse multipath channel model}
We consider the narrow-band mmWave clustered channel model in \cite{Hur2013,Ayach2014,Alkhateeb2014}, where the  matrix  channel $\mathbf{H}$ is assumed to be a sum of the contributions of $N_\text{cl}$ scattering clusters, each of which contribute $N_\text{ray}$ propagation paths. 
The physical channel model $\mathbf{H}\in\mathbb{C}^{N_\text{r}\times N_\text{t}}$ can be expressed as
\begin{equation}
	\begin{aligned}
			\bH= & \sqrt{ \frac{N_\text{t} N_\text{r}} {N_\text{cl} N_\text{ray}}} \sum_{i=1}^{N_\text{cl}} \sum_{\ell=1}^{N_\text{ray}} \beta_{i,\ell}\ba_\text{MS} (\phi_{i,\ell}) \ba_\text{BS}^*(\alpha_{i,\ell}),
\label{eq:geo_ch}
	\end{aligned}
\end{equation}
where  $\beta_{i,\ell}$ is the complex gain of the $\ell^\text{th}$ ray in the $i^\text{th}$ cluster, whereas $\ba_\text{BS} (\alpha_{i,\ell})$ and $\ba_\text{MS}(\phi_{i,\ell})$ are the antenna array response vectors at the transmitter and receiver evaluated at the $\ell^\text{th}$ path $i^\text{th}$ cluster  azimuth angles of departure or arrival. 
The multipath channel model \eqref{eq:geo_ch} can be written in a more compact way as
\begin{equation}
\mathbf{H}=\mathbf{A}_\text{MS}\mathbf{H}_b\mathbf{A}^*_\text{MS},
\end{equation}
where $\mathbf{A}_\text{MS}\in\mathbb{C}^{N_\text{r} \times N_\text{cl}N_\text{ray}}$ and $\mathbf{A}_\text{BS}\in\mathbb{C}^{N_\text{t} \times N_\text{cl}N_\text{ray}}$ contain the array response vectors in the directions $\phi_{i,\ell}$ and $\alpha_{i,\ell}$, and $\mathbf{H}_b$ is a diagonal matrix with the associated path gains $\beta_{i,\ell}$.
MmWave channels are expected to have limited scattering \cite{Rappaport2013}, therefore, a small number of propagation paths $K=N_\text{cl}N_\text{ray}$ is assumed.

The highly directional nature of propagation and the high dimensionality of MIMO channels at mmWave frequencies makes the beamspace representation of MIMO systems a natural choice \cite{mmWaveTutorial2015}.
Assuming that the AoAs and AoDs $\phi,\alpha$ are taken from uniform fine grids of $G_\text{r}$ and $G_\text{t}$ points in $[-\frac{\pi}{2},\frac{\pi}{2})$, we define two dictionary matrices $\mathbf{A}_\text{MSD}=[\ba_\text{MS}(\phi_1) \hdots \ba_\text{MS} (\phi_{G_\text{r})}]$ and $\mathbf{A}_\text{BSD}=[\ba_\text{BS}(\alpha_1) \hdots \ba_\text{BS}(\alpha_{G_\text{r})}]$, with the associated array response vector in these directions.
Neglecting the grid quantization error, we can represent $\mathbf{H}$ in terms of a K-sparse matrix $\mathbf{H}_v\in \mathbb{C}^{G_\text{r} \times G_\text{r}}$ containing the path gains of the quantized angles
\begin{equation}
\mathbf{H}=\mathbf{A}_\text{MSD}\mathbf{H}_v\mathbf{A}_\text{BSD}^*.
	\label{eq:exvirtualChannelModel}
\end{equation}
This representation provides a discretized approximation of the channel response that reduces the task of estimating $\mathbf{H}$ to that of detecting some non zero coefficients in the virtual channel matrix $\mathbf{H}_v$ \cite{Bajwa2010}.
Vectorizing the channel matrix we have
\begin{equation}
\text{vec}(\mathbf{H}) = (\bar{\mathbf{A}}_\text{BSD} \otimes \mathbf{A}_\text{MSD} ) \mathbf{x},
\end{equation}
where $\mathbf{x}=\text{vec}(\mathbf{H}_v)$ is a $G_\text{t}G_\text{r} \times 1$ sparse vector with $K$ non zero entries.
We define the $N_\text{t}N_\text{r} \times G_\text{t}G_\text{r}$ dictionary matrix of the channel $\mathbf{\Psi}=(\bar{\mathbf{A}}_\text{BSD} \otimes \mathbf{A}_\text{MSD} )$.
Each column of $\mathbf{\Psi}$ is of the form $\bar{\mathbf{a}}_\text{BS}(\alpha) \otimes \mathbf{a}_\text{MS}(\phi)$.

Assuming uniform spaced linear arrays with inter-element spacing equal to half wavelength, $\lambda/2$, and the quantized AoAs and AoDs are taken from uniform fine grids of resolution $G_\text{r}=N_\text{r}$ and $G_\text{t}=N_\text{t}$ points, the array response matrices result in unitary DFT matrices $\mathbf{A}_\text{MSD}\in \mathbb{C}^{N_\text{r} \times N_\text{r}}$ and $\mathbf{A}_\text{BSD}\in \mathbb{C}^{N_\text{t} \times N_\text{t}}$.
This case is known as the virtual channel model \cite{Sayeed2002}, which corresponds to beamforming in fixed directions determined by the resolution of the arrays.

\subsection{Compressed channel sensing}

In this section we present a training-based channel estimation approach. 
The mmWave channel estimation is formulated as a sparse reconstruction problem \cite{Alkhateeb2015I,ITA2015}.
The BS and MS sense the channel in open loop with a sequence of training precoders and combiners.
The reconstruction is done applying a standard greedy recovery algorithm.

Consider the mmWave system and channel model described in the previous section.
Let us assume first that the BS and MS employ only one RF chain for the training step.
At time instant $n$, the BS uses a training beamformer $\mathbf{p}_n$ to transmit a symbol $s_n$. 
To simplify the analysis we consider the same symbol $s=1$ in all the transmissions.
The MS applies a training combiner vector $\mathbf{q}_n$ to the received signal. Then the resulting signal can be written as \cite{Alkhateeb2014}
\begin{align}
\label{sensing}
y_{n} & =\sqrt{\rho}\mathbf{q}_n^*\mathbf{H}\mathbf{p}_ns+\mathbf{q}_n^*\mathbf{n} \\
& =\sqrt{\rho}(\mathbf{p}_n^T \otimes \mathbf{q}_n^*) \text{vec}(\mathbf{H})+\mathbf{q}_n^*\mathbf{n} \\
&=\sqrt{\rho}(\mathbf{p}_n^T \otimes \mathbf{q}_n^*)\mathbf{\Psi}\mathbf{x}+\mathbf{q}_n^*\mathbf{n} .
\end{align}
The BS employs $M$ precoding vectors $[\mathbf{p}_1, \hdots, \mathbf{p}_M]$ and the MS employs $M$ combiners $[\mathbf{q}_1, \hdots, \mathbf{q}_M]$ in successive $M$ snapshots.
Stacking the $M$ measurements in vector form we have
\begin{equation}
\label{case1}
\begin{bmatrix}
y_{1} \\
\vdots \\
y_{M} \\
\end{bmatrix} = \sqrt{\rho}
\begin{bmatrix}
\mathbf{p}_1^T \otimes\mathbf{q}^*_1  \\
\mathbf{p}_2^T \otimes\mathbf{q}^*_2  \\
\vdots \\ 
\mathbf{p}_M^T \otimes\mathbf{q}^*_M,  \\
\end{bmatrix}
\mathbf{\Psi}\mathbf{x}+\mathbf{e} .
\end{equation}
Assuming $\mathbf{q}_i^*\mathbf{q}_i=\gamma$ due to hardware constraints, the combined noise vector is given by $\mathbf{e}=[\mathbf{q}^*_1\mathbf{n}_1, \hdots, \mathbf{q}^*_M\mathbf{n}_M]^T$ with covariance $\mathbb{E}[\mathbf{e}\mathbf{e}^*]=\sigma_n^2\gamma\mathbf{I}_{M}$.

The generalization of the approach to many RF chains is as follows.
The BS employs a sequence of $M_\text{t}$ beamforming vectors $\mathbf{P}=[\mathbf{p}_1, \hdots, \mathbf{p}_{M_\text{t}}]$ in successive $M_\text{t}$ time instants.
If the MS utilizes the available $L_\text{r}$ RF chains, it can apply simultaneously $L_\text{r}$ combiners $\mathbf{Q}_n=[\mathbf{q}_1^n, \hdots, \mathbf{q}_{L_\text{r}}^n] \in \mathbb{C}^{N_\text{r} \times L_\text{r}}$ per snapshot to the received signal.
In this way, $L_\text{r}$ channel measurements are produced per time instant.
The sequence of training combiners is defined as $\mathbf{Q}=[\mathbf{Q}_1, \hdots, \mathbf{Q}_{M_\text{t}}] \in \mathbb{C}^{N_\text{r} \times M_\text{t}L_\text{r}}$.
After $M_\text{t}$ snapshots and stacking the $M=M_\text{t}L_\text{r}$ measurements in vector form we have
\begin{equation}
\label{case3}
\begin{bmatrix}
y_{1} \\
\vdots \\
y_{M} \\
\end{bmatrix} = \sqrt{\rho}
\begin{bmatrix}
\mathbf{p}_1^T \otimes\mathbf{Q}^*_1  \\
\mathbf{p}_2^T \otimes\mathbf{Q}^*_2  \\
\vdots \\ 
\mathbf{p}_{M_\text{t}}^T \otimes\mathbf{Q}^*_{M_\text{t}}  \\
\end{bmatrix}
\mathbf{\Psi}\mathbf{x}+\mathbf{e} .
\end{equation}
The noise vector is given by $\mathbf{e}=[\mathbf{Q}^*_1\mathbf{n}_1 \hdots \mathbf{Q}^*_{M_\text{t}}\mathbf{n}_{M_\text{t}}]^T$, with block diagonal covariance matrix $\mathbb{E}[\mathbf{e}\mathbf{e}^*]=\sigma_n^2 \text{diag}(\mathbf{Q}_1^*\mathbf{Q}_1, \hdots, \mathbf{Q}_{M_\text{t}}^*\mathbf{Q}_{M_\text{t}})$.
In general the noise is correlated after combining, except for A5, where $\mathbf{Q}_i^*\mathbf{Q}_i=\mathbf{I}_{L_\text{r}}$ and the noise remains uncorrelated.

Notice that the number of compressed measurements per snapshot depends on the number of available RF chains at the MS but not on the number of RF chains at the BS, since the receive signals are combined at the MS.
The higher the number of RF chains used simultaneously in reception, the higher the number of measurements per snapshot.

\subsection{Channel reconstruction and recovery guarantees}

Since mmWave channels exhibit a sparse-scattering structure, it is possible to leverage theory and algorithms developed for sparse recovery to improve channel estimation results.

Equations $(\ref{case1})$ and $(\ref{case3})$ can be seen as a single measurement vector (SMV) compressive sensing model with the unknown $\mathbf{x}$, a $K$-sparse vector.
Estimating the support of $\mathbf{x}$ is equivalent to estimating the direction of the AoA and AoD of the channel paths, while the values of these non-zero entries correspond to the path gains.
The reconstruction of the channel is formulated as a non-convex combinatorial problem
\begin{equation}
\label{SMV}
	\underset{\mathbf{x}}{\text{minimize}} \  \| \mathbf{x} \|_0 \;\;\; \text{subject to} \;\;\; \| \mathbf{y}-\mathbf{A}\mathbf{x}\|_ 2 \leq \sigma ,
\end{equation}
where the matrix $\mathbf{A}=\mathbf{\Phi}\mathbf{\Psi} \in \mathbb{C}^{ M \times N}$, with $N=G_\text{r}G_\text{t}$.
A variety of Matching Pursuit (MP) algorithms have been proposed for obtaining an approximate solution in polynomial complexity.
We consider Orthogonal Matching Pursuit (OMP) for its simplicity and fast implementation.
OMP has been used for compressed sensing channel estimation \cite{Taubock2010}.

The problem of recovering a high-dimensional sparse signal based on a small number of linear measurements, possibly corrupted by noise, has been extensively studied in compressed sensing.
Recovery guarantees can be obtained providing that the matrix $\mathbf{A}$ has low coherence.
The mutual coherence of a given matrix $\mathbf{A}$ is the largest absolute normalized inner product between different columns from $\mathbf{A}$. 
Suppose $\mathbf{A}=[\mathbf{a}_1 \; \mathbf{a}_2 \; ... \; \mathbf{a}_N]$, the mutual coherence of $\mathbf{A}$ is given by
\begin{equation}
\mu (\mathbf{A}) = \max_{k,j;\ k \neq j} \frac{|\mathbf{a}_k^*\mathbf{a}_j|}{\|\mathbf{a}_k\|_2 \|\mathbf{a}_j\|_2}.
\end{equation}
OMP support recovery guarantee has been stated in the noiseless case in \cite{Tropp2004}.
It was shown that $\mu(\mathbf{A})<\frac{1}{2K-1}$ is a sufficient condition for recovering a $K$-sparse $\mathbf{x}$ exactly in the noiseless case.
This condition is in fact sharp \cite{Cai2010}.
In \cite{Welch1974}, the lower bound of the achievable coherence of a matrix $\mathbf{A}$ of size $M \times N$, with $M \leq N$, is given by the Welch bound
\begin{equation}
\mu(\mathbf{A}) \geq \sqrt{\frac{N-M}{M(N-1)}}.
\label{eq:WB}
\end{equation}
For a given coherence $\mu(\mathbf{A})$, it is possible perfectly to recover a $K$-sparse vector provided $K < \left( \mu(\mathbf{A})^{-1} +1 \right) /2$. 
Therefore, in the noiseless case, the minimum number of measurements $M$ needed to guarantee the recovery of $K$-sparse vectors is given by
\begin{equation}
\label{inequality}
\sqrt{\frac{N-M}{M(N-1)}} < \frac{1}{2K-1}.
\end{equation}

The OMP recovery guarantees in the presence of noise were considered in \cite{Cai2011}.
They require the noise power to be lower than the norm of the non-zero entries in the unknown $\mathbf{x}$. This means that we can guarantee that OMP will perform well if the SNR is not too low.
 
Alternatively, recovery guarantees are provided if $\mathbf{A}$ satisfies the restricted isometry property (RIP).
A matrix $\mathbf{A} \in \mathbb{C}^{M \times N}$ is said to posses the RIP with isometry constant $\delta$ if
\begin{equation}
	(1-\delta)\|\mathbf{x}\|_2^2 \leq \| \mathbf{Ax} \|_2^2 \leq (1+\delta)\| \mathbf{x} \|_2^2 \text{ for all } K\text{-sparse } \mathbf{x},
\end{equation}
and $\delta$ is small enough for sufficiently large $K$. 
The RIP guarantees uniform recovery of $K$-sparse vectors and moreover the reconstructions (by various approximation algorithms) are stable even when sparsity is replaced by compressibility and it is robust in the presence of noise. 
Several classes of matrices that obey RIP, like Gaussian and Rademacher/Bernoulli, are easy to construct \cite{Candes2006, CandesRombergTao_Stable:2006}.

At low SNR, the number of measurement $M$ needs to increase to ensure the quality of the estimate. In fact, the number of measurements is very sensitive to the sparsity of the virtual channel matrix and the SNR. When the sparsity is low,  only a few measurements ($M \ll N$) are needed for good recovery as predicted by the CS theory \cite{Candes2006}.
When the SNR is very low or the sparsity is high, the number of measurement needed is comparable or even higher than the dimension of $\mathbf{x}$.


There are several stopping criteria for OMP that can be implemented with minimal cost. 
A natural way is to stop the algorithm when the sparsity order is reached.
Usually the sparsity order is itself subject to estimation and may not be available. 
Therefore, stopping criteria based on the power of the residual error are often used.
Let $\mathbf{r}^k$ the residual error at iteration $k$, we stop the algorithm when the energy in the residual is smaller than a given threshold $\| \mathbf{r}^k \|_2 \leq \epsilon$. 
A reasonable choice for the stopping threshold is $\epsilon = \mathbb{E}[\mathbf{e}^*\mathbf{e}]$, the noise power after combining. This is the choice made in this manuscript.

\subsection{Training sequences and mutual coherence}

The matrix $\mathbf{A}=\mathbf{\Phi\Psi}$ plays a key role in establishing recovery guarantees for the compressed sensing channel estimation problem.
In this section we design training sequences of precoding/combining vectors that define the sensing matrix $\mathbf{\Phi}$ and provide low coherence.
$\mathbf{\Phi}$ is hardware dependent, therefore, we have to take into account the restrictions imposed by the specific architecture. 
For instance, the phase shifter architecture A1 constrains the precoding/combining vectors $\mathbf{p}$, and $\mathbf{q}$ to have unit norm entries, while a switching architecture A5 constrains the combining/precoding vectors to have exactly a one and zeros elsewhere. In this paper we describe two ways of constructing the sensing matrix, one using the same combiner sequence for all precoders and another using different combiners for each precoder.

\subsubsection{Single combiner matrix (SC)}
We start the analysis considering a more restrictive structure of the sensing matrix.
Consider that the MS applies a fixed sequence of combiners $\mathbf{Q}=[\mathbf{q}_1, \hdots,\mathbf{q}_{M_\text{r}}] \in \mathbb{C}^{N_\text{r} \times M_\text{r}}$ for each precoding vector in $\mathbf{P}=[\mathbf{p}_1,\hdots,\mathbf{p}_{M_\text{t}}]\in \mathbb{C}^{N_\text{t} \times M_\text{t}}$
\begin{equation}
\mathbf{\Phi}=
\begin{bmatrix}
\mathbf{p}_1^T \otimes\mathbf{Q}^*  \\
\mathbf{p}_2^T \otimes\mathbf{Q}^* \\
\vdots \\ 
\mathbf{p}_{M_\text{t}}^T \otimes \mathbf{Q}^*  \\
\end{bmatrix},
\end{equation}
which can be written in a compact way as $\mathbf{\Phi}=(\mathbf{P}^T \otimes \mathbf{Q}^*) \in \mathbb{C}^{M \times N_\text{t}N_\text{r}}$, with $M=M_{\text{t}}M_{\text{r}}$ the number of measurements collected in $M/L_\text{r}$ training steps.
That means that the BS will apply each precoder $\mathbf{p}_i$ during $M_\text{r}/L_\text{r}$ consecutive snapshots, while the MS will apply periodically the same sequence of $M_\text{r}$ combiners in $\mathbf{Q}$.
In this case
\begin{align}
\mathbf{A} &=(\mathbf{P}^T\otimes\mathbf{Q}^*)(\bar{\mathbf{A}}_\text{BSD}\otimes \mathbf{A}_\text{MSD}) \\
&=(\mathbf{P}^T\bar{\mathbf{A}}_\text{BSD}) \otimes (\mathbf{Q}^* \mathbf{A}_\text{MSD}).
\end{align}
Then the coherence of the Kronecker product of two matrices is given by \cite{Jokar2009}:
\begin{equation}
	\mu(\mathbf{A} ) = \max \{ \mu( \mathbf{P}^T \bar{\mathbf{A}}_\text{BSD}  ), \mu( \mathbf{Q}^* \mathbf{A}_\text{MSD} ) \},
\label{eq:MuforKronecker}
\end{equation}
with $\mathbf{P}^T \bar{\mathbf{A}}_\text{BSD} \in \mathbb{C}^{M_\text{t} \times G_\text{t}}$ and $\mathbf{Q}^* \mathbf{A}_\text{MSD} \in \mathbb{C}^{M_\text{r} \times G_\text{r}}$. This allows for the decoupling of the Kronecker product to treat separately the mutual coherence of the two multiplicands. 
In this way the BS and MS can independently design the sequence of precoder/combiners vectors.
This results in a benefit in multi-user environments where each MS can have different architectures and number of antennas.

\begin{figure}[!t]
  \centering
  \centerline{\includegraphics[width=0.5\linewidth]{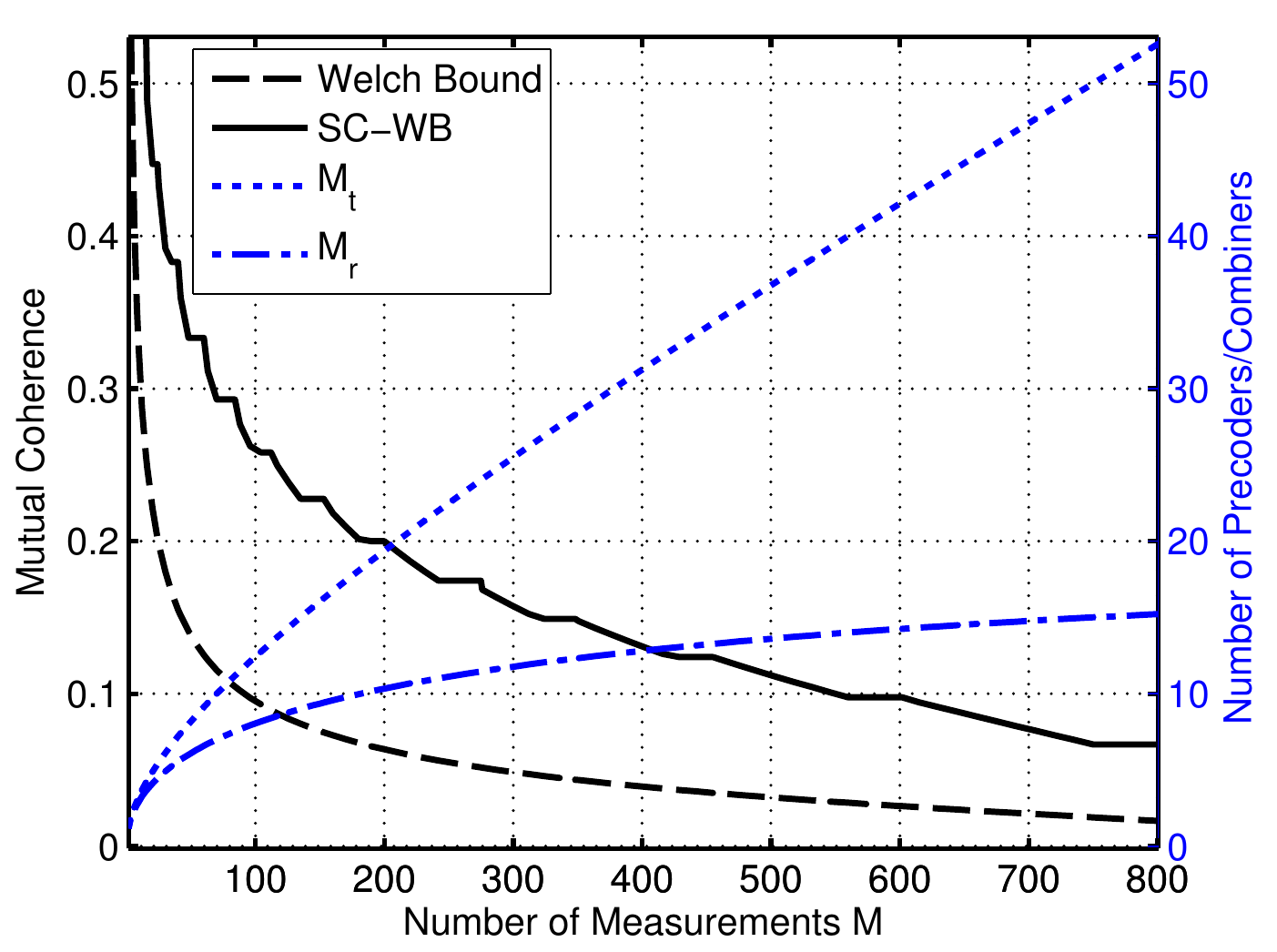}}
  \caption{In blue, the optimum $M_\text{r},M_\text{r}$ that minimizes the coherence of the Kronecker product for a total number of measurements $M=M_\text{t}M_\text{r}$, with $N_\text{t}=G_\text{t}=64$ and $N_\text{r}=G_\text{r}=16$. In black, the single combiner Welch bound (SC-WB) and Welch bound of a general matrix of the same size $M\times G_\text{t}G_\text{r}$.}
\label{fig:coherenceFigure2}
\end{figure}

Given the total number of measurements $M=M_\text{t}M_\text{r}$ and spatial frequency resolutions $G_\text{t} (\approx N_\text{t})$ and $G_\text{r}(\approx N_\text{r})$, one question is how to choose the number of training beamformers $M_\text{t}$ and combiners $M_\text{r}$ to minimize the coherence of $\mathbf{A}$? We can provide a appropriate result based on the Welch bound (WB) -- the minimum mutual coherence that can be theoretically achieved for the dimensions of $\mathbf{A}$ given in \eqref{eq:WB}.
Since both the Welch bounds of the operands $\mathbf{P}^T \bar{\mathbf{A}}_\text{BSD}$ and $\mathbf{Q}^* \mathbf{A}_\text{MSD}$ are decreasing functions of $M_\text{t}$ (or $M_\text{r}$), then $\mu(\mathbf{A})$ is reduced when the Welch bounds of the operands in the Kronecker product are equal.
This equation can be solved using \eqref{eq:WB} and writing it in terms of one unknown $M_\text{t}$ (or $M_\text{r}$).
Therefore, the optimum number of training beamformers/combiners, from a coherence perspective, is given by 
\begin{equation}
\begin{aligned}
M_\text{t} &= -\frac{1}{2}(MG_\text{r}-G_\text{t}M-( 4MG_\text{r}^2G_\text{t}^2+M^2G_\text{r}^2 \\
&-2M^2G_\text{r}G_\text{t}+M^2G_\text{t}^2-4MG_\text{r}^2G_\text{t} \\
&-4MG_\text{r}G_\text{t}^2 +4MG_\text{r}G_\text{t})^{1/2})/(G_\text{r}(G_\text{t}-1)),\\
M_\text{r} &= M/M_\text{t}.
\end{aligned}
	\label{eq:optimumTraining}
\end{equation}
This optimum point exists only for some particular cases of $M_\text{t}$ and $M_\text{r}$, i.e. when both are integer values.
We define the single combiner Welch bound (SC-WB) as the minimum Welch bound of $\mathbf{A}$ for a given number of measurements $M$.
Figure \ref{fig:coherenceFigure2} shows the optimum number of training precoders/combiners $M_\text{t}$ and $M_\text{r}$ that minimize the Welch bound of $\mathbf{A}$ with $G_\text{t}=N_\text{t}=64$ and $G_\text{r}=N_\text{r}=16$, in terms of the total number of measurements $M$.
The SC-WB is shown together with the Welch bound of a general matrix of the same size $M \times G_\text{r}G_\text{t}$.
Notice that the SC-WB is significantly worse than the general Welch bound.

Given $M_\text{t}$ and $M_\text{r}$, we consider the problem of designing training sequences of precoders/combiners $\mathbf{P}$ and $\mathbf{Q}$ to minimize the coherence of $\mathbf{A}_1$ and $\mathbf{A}_2$, conforming with the hardware constraints of the specific hybrid architecture deployed at the BS and MS.
The solutions we provide in this section are based on the virtual channel model since the extended representation using more general overcomplete dictionaries does not lend itself to easy coherence analysis.
We focus on the combining sequence design, as the precoding problem is reciprocal. 

When using switches considering the hybrid architecture A5, $\mathbf{Q}$ is a selection matrix with exactly a ``1" per column.
The problem of selecting rows of the Fourier matrix, $\mathbf{Q}^*\mathbf{A}_{\text{MS}}$ such that the mutual coherence of the resulting matrix is minimized is well studied. 
In some cases, i.e. choices of dimensions, combinatorial sequences based on difference sets \cite{Xia2005} are optimal and achieve the Welch bound.
However, difference sets for typical dimensions of interest in our problem rarely exist.
Alternatively, combinatorial sequences based on almost difference sets \cite{Nowak2014}, can be found for a more general range of dimensions providing near optimal coherence. 
In any case, numerical algorithms to create highly incoherent frames can be applied \cite{Xu2015}\cite{OurPaper2015}.
The same designs can be used for the more general hybrid architecture A4 based on switches.

When using phase shifters A1, $\mathbf{Q}$ is a unital matrix, i.e. a matrix with unit magnitude entries.
To the best of the authors knowledge, there are no optimal designs that achieve the Welch bound with this special structure of the matrix.
Deterministic sequences providing incoherent matrices can be built with the same algorithm \cite{OurPaper2015} since the entries in $\mathbf{Q}^*\mathbf{A}_\text{MS}$ are unit magnitude.

As an alternative to deterministic designs, and according to compressed sensing theory \cite{Bajwa2010}\cite{Ramasamy2012}, random measurements from some classes of distributions (e.g. Rademacher/Bernoulli) provide incoherent matrices satisfying the RIP with high probability. 
Using switches, uniform random selection of rows from a Fourier matrix, i.e. random binary $\mathbf{Q}$ with exactly a ``1" per column, provide equivalent matrices satisfying the RIP \cite{Foucart2013}.
Considering phase shifters A1 and A3, the results in \cite{Bajwa2010}\cite{Ramasamy2012} suggest the use of random sequences $\mathbf{Q}$ with the non zero elements chosen i.i.d. from a discrete uniform distribution taking values $\{ \pm 1, \pm \mathrm{j}\}$. 


\subsubsection{Multiple combiners (MC)}
In the general case, the measurement matrix is given by 
\eqref{case3}
\begin{equation}
\label{LrRF}
\mathbf{\Phi}=
\begin{bmatrix}
\mathbf{p}_1^T \otimes \mathbf{Q}_1^*  \\
\mathbf{p}_2^T \otimes \mathbf{Q}_2^* \\
\vdots \\ 
\mathbf{p}_{M_\text{t}}^T \otimes \mathbf{Q}_{M_\text{t}}^*  \\
\end{bmatrix}.
\end{equation}

\noindent
Analyzing the coherence $\mathbf{A}=\mathbf{\Phi}\mathbf{\Psi}$, with $\mathbf{\Phi}$ given by \eqref{LrRF} is more challenging.
The minimum bound on the coherence with this model is now expected to be lower than for the single combiner case \eqref{eq:MuforKronecker}. This is because we introduce in $\mathbf{\Phi}$ extra degrees of freedom with the different $M_\text{t}$ combiners.

Under the virtual channel model, the dictionary matrix $\mathbf{\Psi}=(\bar{\mathbf{A}}_\text{BS} \otimes \mathbf{A}_\text{MS})$ is a Fourier matrix associated with a 2-dimensional Fourier transform.
Therefore, we suggest the use of pseudorandom sequences or preceders/combiners with the following distributions.
Using phase shifters, A1 and A3, vectors $\mathbf{p}$ and $\mathbf{q}$ with the non zero entries i.i.d. from a uniform distribution with values $\{\pm 1, \pm \mathrm{j}\}$.
Using switches, A5 and A4, binary vectors $\mathbf{p}$ and $\mathbf{q}$ with zeros and only a ``1" with its position uniformly random distributed.

\begin{figure*}[!t]
\centering
\subfloat[Mutual coherence of the equivalent measurement matrix $\mathbf{\Psi}$ designed with pseudorandom sequences for different architectures and considering one RF chain at the MS.]{\includegraphics[width=0.5\linewidth]{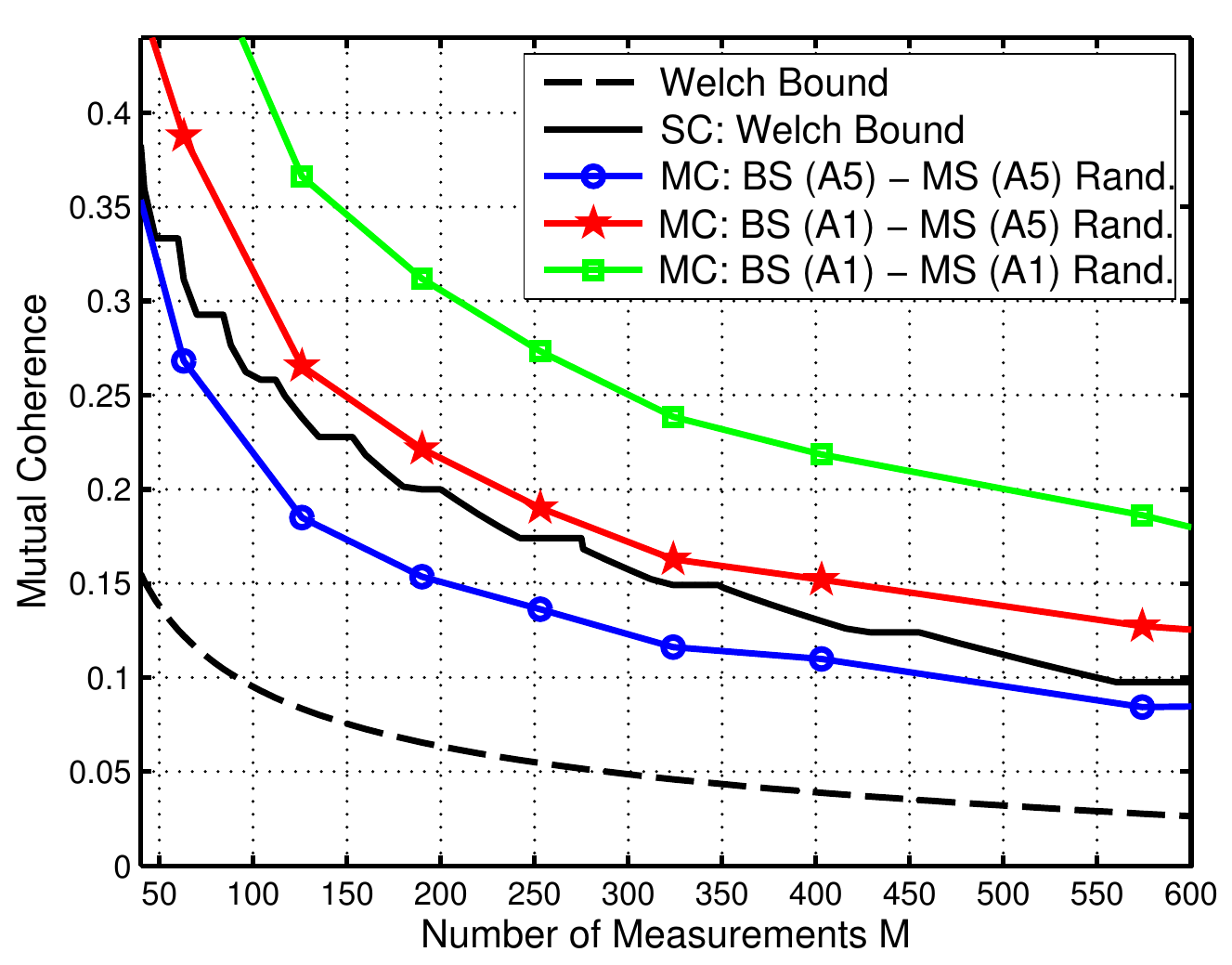}%
\label{fig:Rand}}
\; \;
\subfloat[Mutual coherence of the equivalent measurement matrix $\mathbf{\Psi}$ designed with \cite{OurPaper2015} for different architectures and considering one RF chain at the MS.]{\includegraphics[width=0.5\linewidth]{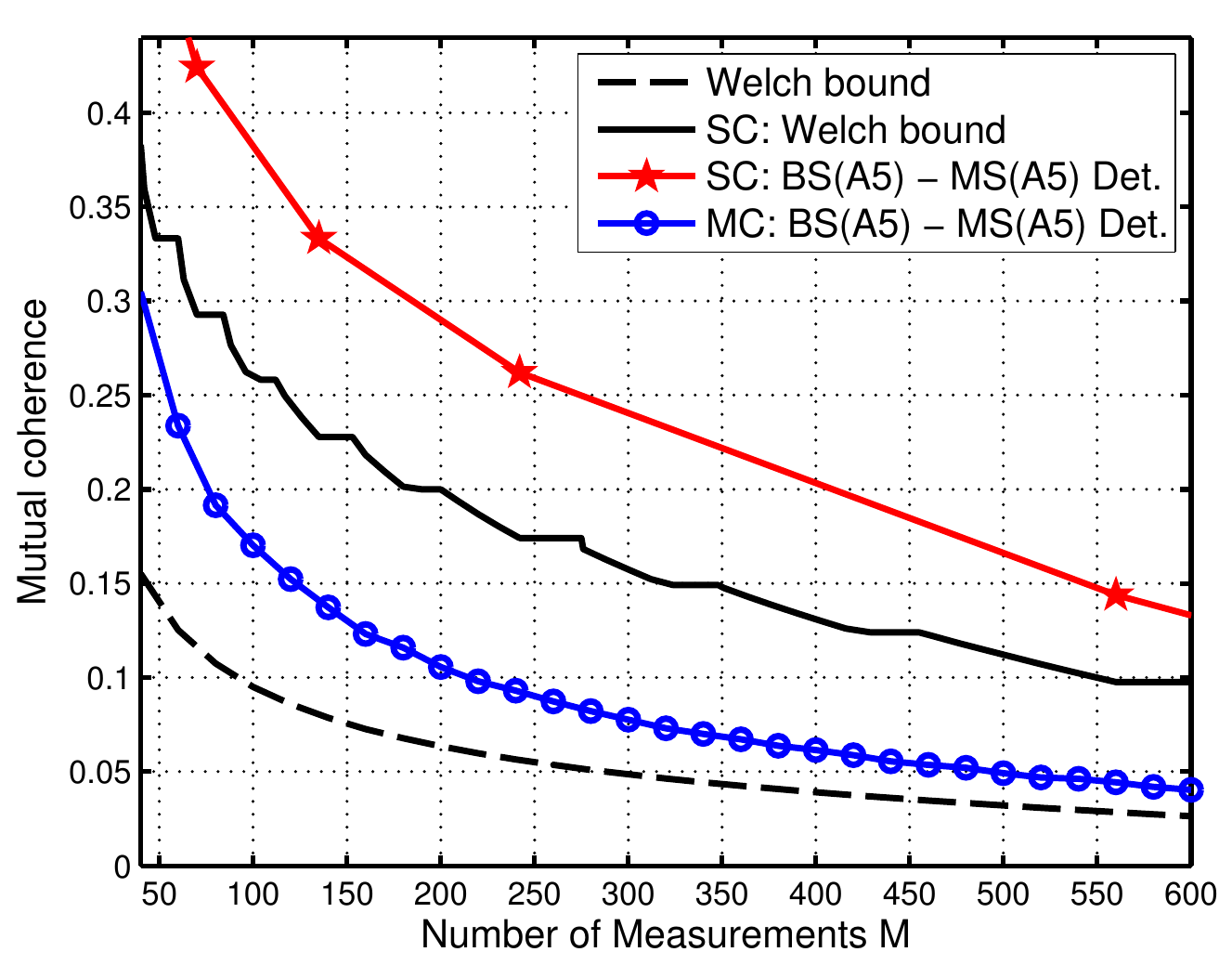}%
\label{fig:Det}}
\caption{A comparison of mutual coherence for pseudorandom and deterministic designs.}
\label{fig:RandAndDet}
\end{figure*}

Figure \ref{fig:RandAndDet}\protect\subref{fig:Rand} shows the mutual coherence $\mu(\mathbf{A})$ with $G_\text{t}=N_\text{t}=64$ and $G_\text{r}=N_\text{r}=16$ achieved with the proposed pseudorandom sequences in the MC case for different combinations of hybrid architectures at the BS and MS.
Only one RF chain is considered at the MS, $L_\text{r}=1$.
The curves show the lowest coherence achieved from $100$ random realizations.
The Welch bound and SC-WB bound are also plotted for comparison.
The achieved mutual coherence with pseudorandom sequences in the MC case with switches at BS and MS is lower than the single combiner Welch bound, which shows the superiority of the MC approach. 
Notice that switches, A5-A5, provide better coherence than phase shifters, A1-A1, with a noticeable power consumption reduction.

As an alternative to random training sequences, it is also possible design deterministic measurement matrices to reduce the coherence of $\mathbf{A}$.
In the MC case, the sequence of precoder and combiner vectors appear coupled in the optimization problem.
That means that an optimization problem needs to be solved for each specific combination of number of antennas and RF chains deployed at the BS and MS.

In this multiple combiner case, we analyze a switches (A5-A5) architecture with multiple RF chains. The goal is to create measurement matrices
\begin{equation}
	\mathbf{A} = \begin{bmatrix}
							\mathbf{s}_1^T \bar{\mathbf{A}}_\text{BSD} \otimes \mathbf{S}_1 \mathbf{A}_\text{MSD} \\
							\mathbf{s}_2^T \bar{\mathbf{A}}_\text{BSD} \otimes \mathbf{S}_2 \mathbf{A}_\text{MSD} \\
							\vdots \\
							\mathbf{s}_{M_\text{t}}^T \bar{\mathbf{A}}_\text{BSD} \otimes \mathbf{S}_{M_\text{t}} \mathbf{A}_\text{MSD} \\
							\end{bmatrix},
\end{equation}
that have coherence as low as possible. The vectors $\mathbf{s}_j^T$ of size $1 \times N_\text{t}$ are all zeros except for one entry that takes the value ``1" and the matrices $\mathbf{S}_j$ of size $M_\text{r} \otimes N_\text{r}$ have a ``1" value on each row and zero everywhere else. The final measurement matrix $\mathbf{\Phi}$ is of size $M_\text{t}M_\text{r} \times N_\text{t}N_\text{r}$.

To construct deterministically such matrices with low mutual coherence we deploy a (suboptimal) greedy approach. Notice that the problem reduces to selecting multiple times rows of $\bar{\mathbf{A}}_\text{BSD}$ and $\mathbf{A}_\text{MSD}$ such that the overall coherence is minimized. For each operand at a time, in all the Kronecker products, the algorithm selects  rows iteratively such that the current coherence is maximally reduced. Convergence is attained when a complete sweep of all $\mathbf{s}_j^T$ and $\mathbf{S}_j$ does not improve the coherence.

Figure \ref{fig:RandAndDet}\protect\subref{fig:Det} shows the mutual coherence $\mu(\mathbf{A})$ with $G_\text{t}=N_\text{t}=64$ and $G_\text{r}=16$ achieved with \cite{OurPaper2015} for different architectures and considering one RF chain at the MS.
The architecture based on switches achieves the lowest coherence.
In comparison to Figure \ref{fig:RandAndDet}\protect\subref{fig:Det}, the deterministic designs provide lower coherence than the random sequences, below the SC-WB in most cases.



To summarize, we propose a measurement matrix structure that includes multiple combiners at the receiver allowing therefore lower coherence than the SC case. For the single RF chain, we propose a deterministic method to design low coherence training sequences for the architectures A1 and A5 based on Legendre and maximum length sequences.

\subsection{Least squares estimation} 
Linear least square is a conventional approach for channel estimation in MIMO systems \cite{Karami2007}\cite{Biguesh2006}.
In traditional training-based methods considering architectures with one RF chain per antenna, it is well known that any sequence of training symbols with orthogonal rows of the same norm is optimal in the least-squares sense \cite{Biguesh2006}.
In the hybrid precoding framework,  the RF precoding/combining matrices must be included in the sensing model \eqref{sensing}. 
The channel estimation model is given by 
\begin{align}
\mathbf{y} = 
\sqrt{\rho}\mathbf{\Phi}
\mathbf{h}_v+\mathbf{e},
\end{align}
with $\mathbf{h}_v = \text{vec}(\mathbf{H})$.
The LS estimation yields
\begin{equation}
\label{eq_ls_estimator}
\mathbf{h}_{v_\text{LS}} = (\mathbf{\Phi}^*\mathbf{\Phi})^{-1}\mathbf{\Phi}^*\mathbf{y}.
\end{equation}
This requires the sensing matrix $\mathbf{\Phi}$ to be full column rank.
We can find $\mathbf{\Phi}$ that minimizes the channel estimation error subject to power and hardware constraints, considering that only one RF chain is used at the receiver.
The general sensing matrix \eqref{LrRF} is defined by the sequence of training precoders $\mathbf{P}=[\mathbf{p}_1, \hdots, \mathbf{p}_{M_\text{t}}]$, assuming a total transmit power constraint $\| \mathbf{P}\|_F^2=\mathcal{P}$, and the respective set of combiners $\mathbf{Q}=[\mathbf{q}_1, \hdots, \mathbf{q}_{M_\text{t}}]$ with the hardware constraint $\|\mathbf{q}_i\|_2^2=\gamma$, for $i = 1,\dots,M_\text{t}$.
The aim is to design $\mathbf{\Phi}$ that minimizes the channel estimation error
\begin{align}
\label{ls_error}
\underset{\mathbf{\Phi}}{\text{minimize }} & \mathbb{E}[ \|\mathbf{h}_v-\mathbf{h}_{v_\text{LS}}\|_2^2] \\
\text{subject to } & \|\mathbf{P} \|_F^2 = \mathcal{P},\ \mathbf{q}_i^*\mathbf{q}_i = \gamma.  \nonumber
\end{align}
This problem can be solved using the Lagrange multiplier method and following the derivation in \cite{Biguesh2006}.
Any sensing matrix is optimal if it satisfies
\begin{equation}
\label{eq_opt_condition}
\mathbf{\Phi}^*\mathbf{\Phi}=\frac{\gamma \mathcal{P}}{N_\text{t}N_\text{r}}\mathbf{I}.
\end{equation}
This result in a minimum MSE
\begin{equation}
J_\text{LS}=\frac{\sigma_n^2 N_\text{t}^2N_\text{r}}{\mathcal{P}}.
\end{equation}
Notice that the error increases with the square of the number of transmit antennas.

Until now we have omitted the hardware constraints on $\mathbf{\Phi}$. Now we show how to find implementable sensing matrices with the set of feasible precoders/combiners associated to the particular architecture that fulfill \eqref{eq_opt_condition}.
Consider the restrictive structure of the sensing matrix in the SC case $\mathbf{\Phi}=(\mathbf{P}^T \otimes \mathbf{Q}^*)$, with $\mathbf{P} \in \mathbb{C}^{N_\text{t} \times M_\text{t}}$, $\mathbf{Q} \in \mathbb{C}^{N_\text{r} \times M_\text{r}}$ and $M=M_{\text{t}}M_{\text{r}}$ the total number of measurements, we can write
\begin{align}
\mathbf{\Phi}^*\mathbf{\Phi} &=(\mathbf{P}^* \otimes \mathbf{Q})(\mathbf{P}^T \otimes \mathbf{Q}^*) \\
&=(\mathbf{P}\mathbf{P}^*)^T \otimes (\mathbf{Q}\mathbf{Q}^*).
\end{align}
An equivalent condition to \eqref{eq_opt_condition} is that $\mathbf{P}\mathbf{P}^*=\frac{\mathcal{P}}{N_\text{r}}\mathbf{I}_{N_\text{t}}$ and $\mathbf{Q}\mathbf{Q}^*=\frac{\gamma}{N_\text{r}}\mathbf{I}_{N_\text{r}}$.
We provide the following designs for $\mathbf{Q}(\mathbf{P})$ satisfying these conditions:
\begin{itemize}
\item A1: $\mathbf{Q}$ defined as the first $N_\text{r}$ rows of a Fourier or Hadamard matrix $\mathbf{F} \in \mathbb{C}^{M_\text{r} \times M_{\text{r}}}$.

\item A2: a block diagonal matrix $\mathbf{Q}=\text{diag}(\hat{\mathbf{Q}}_1,\hdots,\hat{\mathbf{Q}}_{L_\text{r}})$, with each block $\hat{\mathbf{Q}}_i$ designed from the first $N_\text{r}/L_\text{r}$ rows of a Fourier or Hadamard matrix $\mathbf{F} \in \mathbb{C}^{M_\text{r}/L_\text{r} \times M_\text{r}/L_\text{r}}$

\item A3 and A5: $\mathbf{Q}=[\mathbf{I_{N_\text{r}}},  \hdots, \mathbf{I_{N_\text{r}}}] \in \mathcal{B}^{N_\text{r} \times M_\text{r}}$, which results from the concatenation of $ M_\text{r}/N_\text{r}$ identity matrices $\mathbf{I}_{N_\text{r}}$.

\item A4 and A6: a block diagonal matrix $\mathbf{Q}=\text{diag}(\hat{\mathbf{Q}}_1,\hdots,\hat{\mathbf{Q}}_{L_\text{r}})$, with each block $\hat{\mathbf{Q}}_i$ defined as the first $N_\text{r}/L_\text{r}$ columns of the identity matrix $\mathbf{I}_{M_\text{r}/L_\text{r}}$.
\end{itemize}

The generalization of the optimal LS condition when many RF chains are used simultaneously at the receiver is not straightforward.
In that case, the covariance of the noise after combining is not a diagonal matrix in general, which complicates the analysis.
For architectures A5 and A6, the optimality condition \eqref{eq_opt_condition} holds since $\mathbf{Q}_i^*\mathbf{Q}_i=\mathbf{I}_{L_{r}}$.



\section{Hybrid combining with switches} \label{sec:combine}
We seek to design hybrid combiners $\mathbf{W}=\mathbf{W}_\text{RF}\mathbf{W}_\text{BB}$ to maximize the spectral efficiency of the MIMO system in Fig.~\ref{fig:hybrid}.
We only focus on the receiver design, using any of the hybrid architectures in Fig.~\ref{fig_sim}.
For this result, we assume 1) perfect channel knowledge (CSI) at the receiver and 2) the transmitter uses an unconstrained optimal baseband precoder. 
We consider the approach in \cite{Ayach2014} where the joint transmitter-receiver maximization of the mutual information is temporarily decoupled and the focus at the receiver falls on designing $\mathbf{W}_\text{RF}$ and $\mathbf{W}_\text{BB}$ to maximize the mutual information,
\begin{equation}
I(\mathbf{H}) = \log_2 \left | \mathbf{I}_{N_\text{t}}+\frac{\text{SNR}}{N_\text{t}}\tilde{\mathbf{H}}^*\mathbf{W}(\mathbf{W}^*\mathbf{W})^{-1}\mathbf{W}^*\tilde{\mathbf{H}} \right |.
\label{eq:capacityHybrid}
\end{equation}
We have denoted the virtual channel $\tilde{\mathbf{H}} = \mathbf{H} \mathbf{F}_\text{opt}$ where $\mathbf{H}$ is the channel matrix and $\mathbf{F}_\text{opt}$ is the unconstrained optimum precoder applied at the transmitter.
Given the singular value decomposition of the channel $\mathbf{H}=\mathbf{U\Sigma V}^*$, the optimum unconstrained combiner $\mathbf{W}_\text{opt}=\mathbf{U}_{N_\text{s}}$ is given by the first $N_\text{s}$ singular vectors  of the unitary matrix $\mathbf{U}$.
However, when the specific hardware constraints associated to the mmWave architectures are considered in the design, maximizing \eqref{eq:capacityHybrid} leads to a intractable constrained optimization problem.

For the architecture A1, a low complexity hybrid precoding/combining solution was proposed in \cite{Ayach2014}.
The algorithm exploits the sparse structure of the mmWave channel and formulates the design as a sparse reconstruction problem solved with a variant of simultaneous orthogonal matching pursuit (SOMP).
Other efficient alternatives based on projections on matrices with unitary entries can be found in \cite{Molisch2005}\cite{spawc2015}. 

For architectures including a subset antenna selection step, i.e. A5 or A6, the problem of designing the optimal combiner is combinatorial. 
The only known exact solution is by exhaustive search, which is time consuming.
In this context, near optimal solutions can be found in \cite{Gharavi-alkhansari2004} or \cite{ITA2015}.
The algorithm in \cite{Gharavi-alkhansari2004} focuses only in the design of the analog switching stage, while \cite{ITA2015} provides a hybrid design including the baseband combiner.
The method in \cite{Gharavi-alkhansari2004} is an incremental successive selection algorithm that at each step activates the antenna that provides the largest increase in capacity.
In the case of antenna selection in subsets A6, the same algorithm can be applied modifying the set of feasible antennas we can select at each iteration.
That is, when one antenna form the subset $i \in[1 \hdots L_\text{r}]$ is activated, we have to remove from the feasible set all the indexes associated to antennas in the subset $i$.
The hybrid antenna selection algorithm in \cite{ITA2015} extends the strategy in \cite{Gharavi-alkhansari2004} to account for the hybrid combiner design.
The procedure is summarized in Algorithm \ref{alg:HAS}.


\begin{algorithm}[hbtp]
\caption{ \textbf{-- Hybrid Antenna Selection}}\label{alg:HAS}
\begin{algorithmic}

\State \textbf{Initialization:} Select the first $N_\text{s}$ rows of $\mathbf{H}$ according to the greedy antenna selection algorithm in \cite{Gharavi-alkhansari2004}. The rows are indexed in the set $\mathcal{S}$ and the channel matrix restricted to this support is denoted by $\mathbf{H}_\mathcal{S}$.

\State \textbf{Iterations }$k = N_\text{s}+1,\dots ,L_\text{r}$:
	\begin{enumerate}
	\setlength{\itemindent}{+.1in}
		\item For each index $j \in \{1,\dots,N_\text{r}\} \backslash \mathcal{S}$:
			\begin{enumerate}
				\item Construct new channel matrix $\mathbf{H}_{\mathcal{S} \cup j}$.
				\item Perform the singular value decomposition of $\mathbf{H}_{\mathcal{S} \cup j}$ and set the baseband combiner $\mathbf{W}_\text{BB}$ as the left singular vectors.
				\item Compute the mutual information \eqref{eq:capacityHybrid} with $\mathbf{H}_{\mathcal{S} \cup j}$ and its associated $\mathbf{W}_\text{BB}$. Save result in variable $C(j)$.
			\end{enumerate}

		\item Choose $j_\text{opt} = \underset{j}{\text{argmax}}\ C(j)$.

		\item Update support $\mathcal{S} = \mathcal{S} \cup j_\text{opt}$.
	\end{enumerate}
\State Set the RF combiner $\mathbf{W}_\text{RF}$ as the first $L_\text{r}$ rows from the identity matrix $\mathbf{I}_S$ indexed in the set $\mathcal{S}$.

\end{algorithmic}
\end{algorithm}

For the structures A2, A3 and A4, the joint hybrid combining design is still an open problem \cite{Han2015}.
Further, exhaustive search is not feasible due to the extreme number of possible combinations that would need to be checked.
Alternatively, in this paper we propose two low computational complexity hybrid combiner designs based on \cite{Ayach2014}.
Assuming that $\mathbf{W}=\mathbf{W}_\text{RF}\mathbf{W}_\text{BB}$ can be made mathematically ``close" to the unconstrained combiner, near optimal combiners that maximize \eqref{eq:capacityHybrid} can be found by minimizing the ``distance" between $\mathbf{W}_\text{RF}\mathbf{W}_\text{BB}$ and $\mathbf{W}_\text{opt}$ \cite{Ayach2012}.
We seek to solve the next constrained optimization problem in terms of the Frobenius norm
\begin{align}
	 \underset{\mathbf{W}_\text{RF},\mathbf{W}_\text{BB}}{\text{minimize }} &  \| \mathbf{W}_\text{opt} -\mathbf{W}_\text{RF}\mathbf{W}_\text{BB}\|_F \nonumber \\
	 \text{subject to }&  \mathbf{W}_\text{RF} \in \mathcal{W}_\text{RF},
\label{eq:fro}
\end{align}
where $\mathcal{W}_\text{RF}$ is the set of feasible RF combiners associated with the mmWave system architecture.
Notice than solving \eqref{eq:fro} is not equivalent to maximizing the spectral efficiency.
Nevertheless, the use of norm-based designs is interesting because of its low computational complexity. Also, it has been shown in \cite{Ayach2012} that minimizing the Frobenius norm bound results from maximizing a bound on the achievable spectral efficiency.

To simplify the problem, we use the same strategy as \cite{Ayach2014}.
Defining a dictionary of feasible combiners $\mathbf{D}=[\mathbf{d}_1, \hdots, \mathbf{d}_{N_\text{d}}]$ of size $N_\text{r} \times N_\text{d}$, we solve the problem
\begin{align}
\label{eq:opt_froBd}
\underset{\tilde{\mathbf{W}}_\text{BB}}{\text{minimize }} & \| \mathbf{W}_\text{opt} - \mathbf{D} \tilde{\mathbf{W}}_\text{BB} \|_F \nonumber \\
 \text{subject to }  & \| \text{diag}(\tilde{\mathbf{W}}_\text{BB} \tilde{\mathbf{W}}_\text{BB}^*) \|_0 = L_\text{r} \nonumber \\
  & \| \mathbf{D} \tilde{\mathbf{W}}_\text{BB} \|_F^2 = N_\text{s},
\end{align}
where $\| \cdot \|_0$ is the $\ell_0$ pseudo-norm accounting for the number of non-zero elements, $\tilde{\mathbf{W}}_\text{BB} \in \mathbb{C}^{N_\text{d} \times N_\text{s}}$ has only $L_\text{r}$ non-zero rows and an energy constraint.
$\mathbf{W}_\text{BB} \in \mathbb{C}^{L_\text{r} \times N_\text{s}}$ is given by the $L_\text{r}$ non-zero rows from $\tilde{\mathbf{W}}_\text{BB}$.
The solution to the problem can be obtained with the simultaneous orthogonal matching pursuit (SOMP) in \cite{Ayach2014}. 
Different dictionary matrices are defined for A2, A3 and A4.

We first solve for the case of architecture A2 with each RF chain connected to a subset of antennas. This architecture can be seen as the combination of $L_\text{r}$ phased arrays of lower dimension, with $N_\text{r}/L_\text{r}$ antennas each. The set of feasible combiners is given by $\mathcal{A}_2$. From this set we create a block diagonal dictionary matrix $\mathbf{D}=\text{diag}(\mathbf{D}_1, \hdots, \mathbf{D}_{L_\text{r}})$, with each block $\mathbf{D}_i=\tilde{\mathbf{D}}$ of size $N_\text{r}/L_\text{r} \times N'_\text{d}$ associated to each phased array. We define $\tilde{\mathbf{D}}=[\tilde{\mathbf{d}}(\theta_1), \hdots, \tilde{\mathbf{d}}(\theta_{N'_\text{d}})]$ a dictionary of steering vectors associated to each phased array with an angular resolution $N'_\text{d}$. The size of the complete dictionary $\mathbf{D}$ is $N_\text{r} \times {N_\text{d}}$ with ${N_\text{d}}=N'_\text{d}L_\text{r}$.

We next propose solutions for the architectures A3 and A4. These architectures provide more flexibility than A5 and A6 at the expenses of increasing the complexity and associated power consumption. The ability of A3 and A4 to select and sum the contribution from a subset of antennas prior feeding the RF chain, contributes to increase the antenna effective area and the array gain with respect to A5 and A6. Unfortunately, combining no co-phased received signals can also cause significant degradation. 
Considering A3, the number of antennas in the selected subset varies from $0$ to $N_\text{r}$, since each switch can be activated or not. Therefore, the phase network is modulated as a binary matrix with ``0" and ``1" entries.
The set of feasible combiners associated with A3 is given by the set $\mathbf{D}=\mathcal{A}_3$.
The dimension of the dictionary $N_\text{d}$ grows exponentially with the number or antennas $2^{N_\text{r}}$.
For a small number of antennas, e.g., $N_\text{r}=8$, the dimension $N_\text{d}=256$ is reasonable.
However, for higher $N_\text{r}$, $N_\text{d}$ rapidly increases and the size of the dictionary becomes intractable.
Therefore, choosing a reduced dictionary of combiners is necessary in some cases. 
This is an open problem that we do not consider in this paper.
Similarly for A4, the set of feasible combiners is given by $\mathcal{A}_4$. 
The architecture imposes the constraint that only one combiner from each block in the dictionary can be selected.

\section{Simulations}\label{sec:results}

\begin{table}[htbp]
  \centering
    \begin{tabular}{|c|c|c|c|c|c|c|c|c|c|}
    \hline
          & \multicolumn{7}{c}{Channel estimation}  \vline         & \multicolumn{2}{c}{Hybrid combining}   \vline \\
    \hline
          & \multicolumn{4}{c}{Method}   \vline & \multicolumn{3}{c}{Training sequences} \vline &   \multicolumn{2}{c|}{}  \\
          \hline
					& LS    & OMP   & Exhaustive Search    & Adaptive CS & Random & \multicolumn{2}{c|}{Deterministic} & SOMP  & HAS \\
					\cline{7-8}
					&   &    &     &  &  & Incoherence & Orthogonal &   & \\
					\hline
    A1    & x     & x     & x     & x     & x     &       & x     & x     &  \\
		\hline
    A2    & x     & x     & x     &       & x     &       & x     & x     &  \\
		\hline
    A3    & x     & x     & x     &       & x     & x     & x     & x     &  \\
		\hline
    A4    & x     & x     & x     &       & x     & x     & x     & x     &  \\
		\hline
    A5    & x     & x     & x     &       & x     & x     & x     &       & x \\
		\hline
    A6    & x     & x     & x     &       & x     & x     & x     &       & x \\
    \hline
    \end{tabular}%
    \caption{Outline of the channel estimation and hybrid combining methods used for each proposed architecture.}
  \label{tab:summary}%
\end{table}%

In this section, we evaluate the performance of the proposed algorithms for channel estimation and combiner optimization. First, we provide results on mean squared error of the proposed adaptive channel estimator. Second, we provide results on achievable spectral efficiency. We evaluate the trade-off achievable spectral efficiency-power consumption with each hybrid architecture considering the power consumption model in Section~\ref{sec:powermodel}.Table \ref{tab:summary} contains a summary of the channel estimation and hybrid combining algorithms that can be used for each proposed architecture.

\subsection{Channel estimation}

First, we evaluate the channel estimation performance using the proposed compressed sensing based channel estimation approach and the least squares solution.
We use the normalized mean square error as the performance metric NMSE $=\frac{\| \mathbf{H}-\hat{\mathbf{H}} \|_F^2}{\|\mathbf{H}\|_F^2}$.
We consider a single user downlink channel with $N_\text{t}=64$ transmit antennas and $L_\text{t}=8$ RF chains at the BS and $N_\text{r}=16$ antennas and $L_\text{r}=4$ RF chains at the MS.
To evaluate the impact of the quantization error in the AoAs/AoDs estimation, two different channel models are considered. 
First, a simple channel with $N_\text{cl}=4$ clusters and $N_\text{ray}=1$ with the AoAs and AoDs uniformly distributed on quantized grids with resolutions $G_\text{t}$ and $G_\text{r}$. 
In this case, the channel matrix $\mathbf{H}_v$ has $K=4$ non-zero entries.
Second, a more realistic cluster channel model with $N_\text{cl}=4$ clusters and $N_\text{ray}=6$ per cluster with unquantized AoAs and AoDs in $[0,2\pi)$.

In the first experiment, we compare the proposed compressed sensing based approach and least squares with two beam scanning techniques: an exhaustive search training algorithm and the adaptive compressed sensing approach in \cite{Alkhateeb2014}, both assuming a hybrid architecture based on phased shifters.
Although beam training algorithms are used to design implicitly the precoders and combiners, both strategies can be employed to obtain an estimate of the channel explicitly by estimating the $K$ dominant paths and their associated AoAs/AoDs and path gain \cite{Alkhateeb2014}.
Exhaustive search sounds the channel using high resolution beams testing all possible combinations from codebooks of resolutions $G_\text{t},G_\text{r}$.
Therefore, $G_\text{r} \times G_\text{t}$ measurements are needed to estimate the channel, which results in $G_\text{r} \times G_\text{t}/L_\text{r}$ training steps if $L_\text{r}$ RF chains are used simultaneously at the receiver.
The method in \cite{Alkhateeb2014} is a beam training process that involves adaptive measurements over a multiresolution dictionary reducing the training overhead with respect to exhaustive search.
The algorithm estimates the parameters (AoA/AoD and path gain) of one path per iteration after subtracting the contribution of the previously estimated paths.
To estimate the parameters of each path, an adaptive search over the AoA/AoD is performed starting with wide beams in the early stages and narrowing the search based on the estimation outputs in the later stages to focus only on the most promising directions. 
One drawback of the adaptive scheme is the need for a feedback link between the transmitter and receiver.
The required number of training steps to estimate the channel depend on the number of dominant paths $K$.
Considering $K=4$ paths, $L_\text{r}=4$ RF chains, and a possible choice for the grid resolution $G_\text{r}=G_\text{t}=64$ \cite{Alkhateeb2014}, the number of required training steps is $256$ $\left(LK^2[\frac{LK}{L_\text{r}}]\log_L \left( \frac{G_\text{r}}{K} \right) \text{ with } L=2\right)$.
For comparison purpose we consider the same number of training steps and grid resolution for the proposed CS and LS schemes, which results in $256 \times L_\text{r}=1024$ measurements.
In this case, the training sequences for OMP are designed as pseudorandom vectors with entries $\{\pm 1,\pm \mathrm{j}\}$.
For LS, the training sequences $\mathbf{P}$ and $\mathbf{Q}$ are designed to provide orthogonality.

\begin{figure*}[!th]
\centering
\subfloat[Simple channel model $N_\text{cl}=4$, $N_\text{ray}=1$]{\includegraphics[width=0.5\linewidth]{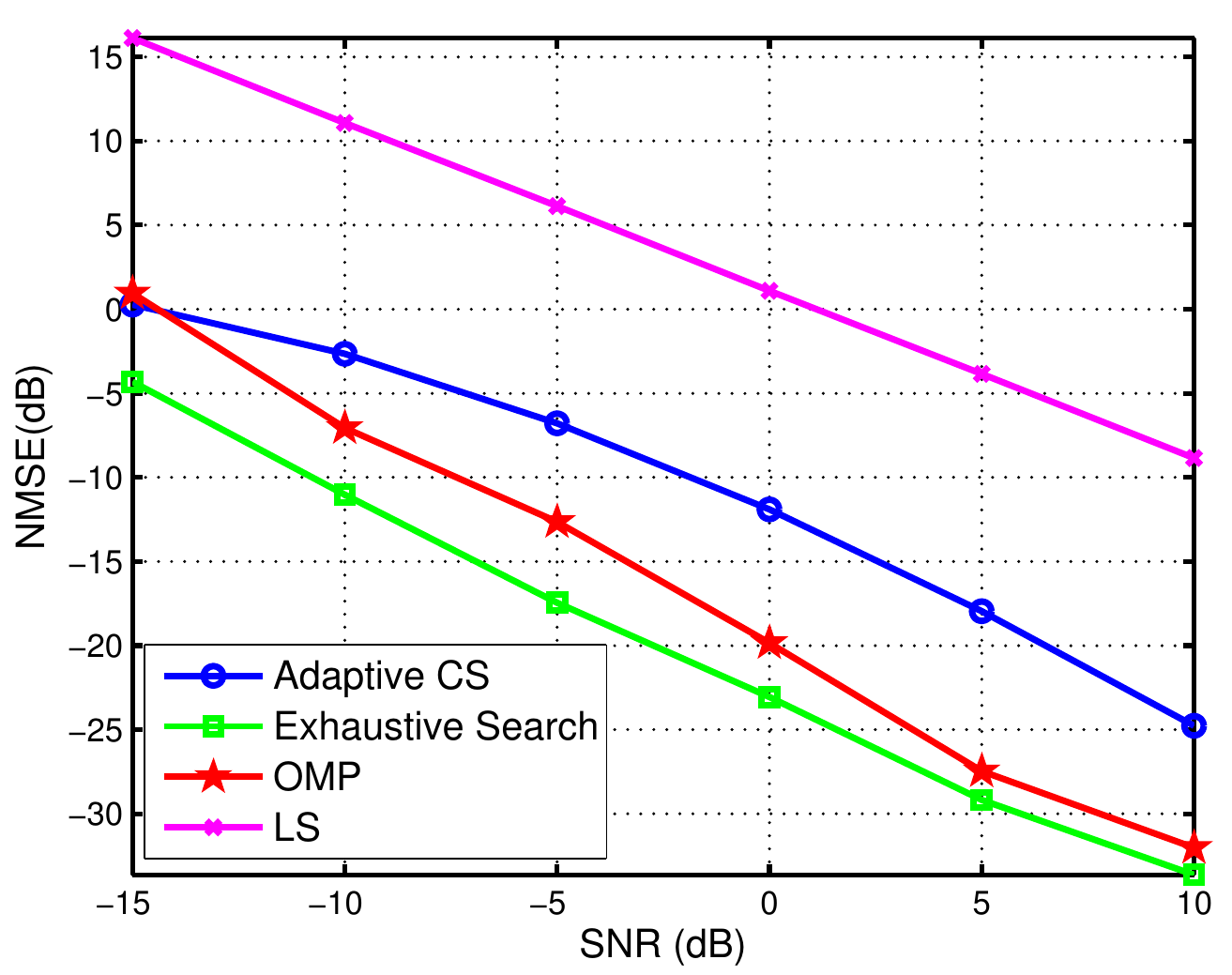}%
\label{fig:NMSE_on}}
\subfloat[Cluster Channel model $N_\text{cl}=4$, $N_\text{ray}=6$]{\includegraphics[width=0.5\linewidth]{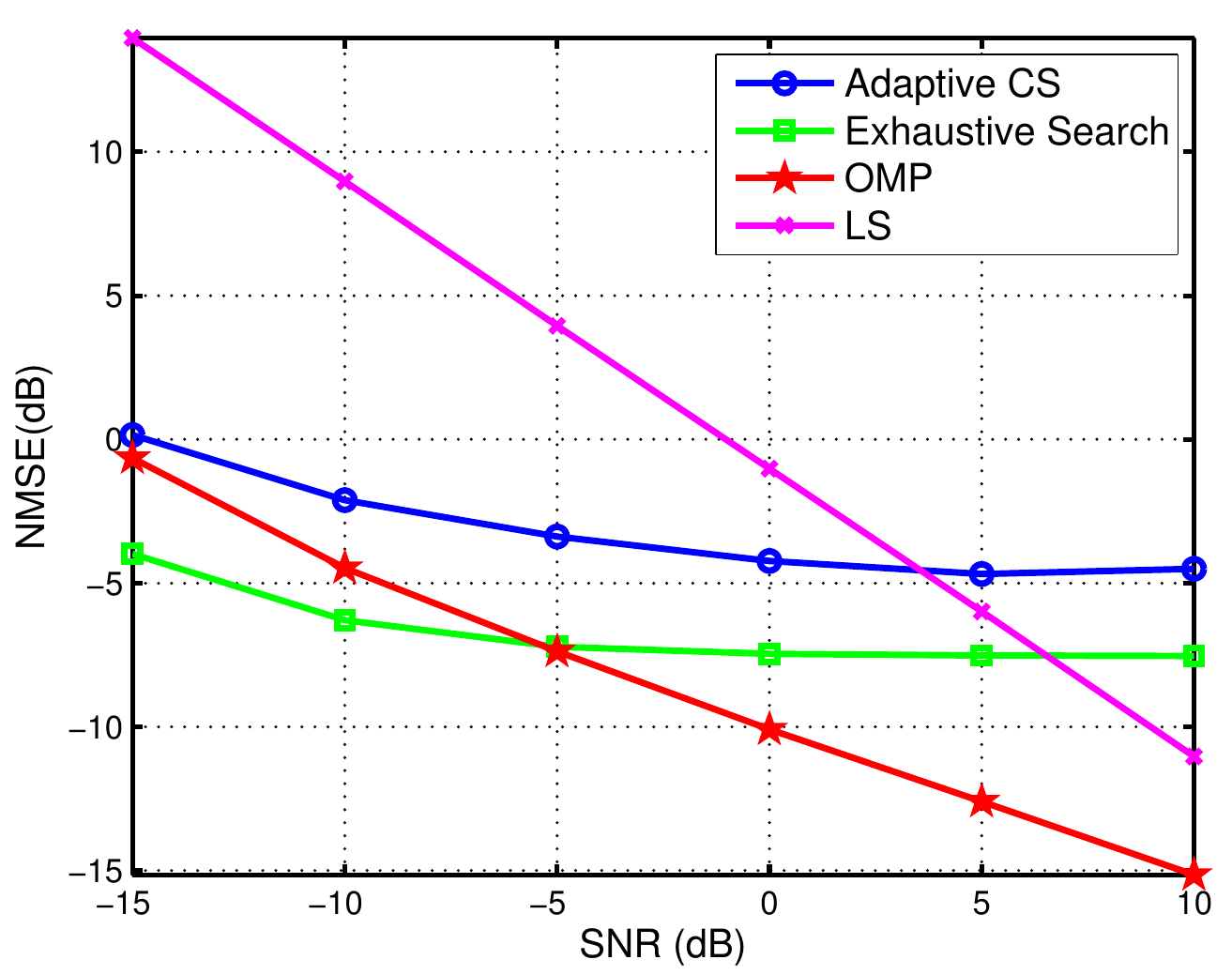}%
\label{fig:NMSE_off}}
\caption{Normalized Estimated Channel MSE vs SNR: $N_\text{t}=64$, $N_\text{r}=16$, $L_\text{t}=8$, $L_\text{r}=4$, and $G_\text{r}=G_\text{t}=64$. Number of training steps $M=256$.}
\label{fig:NMSEvsSNR}
\end{figure*}

Fig. \ref{fig:NMSEvsSNR} shows the NMSE obtained with OMP and LS together with exhaustive search and adaptive CS for different SNR values.
In Fig. \ref{fig:NMSE_on}, OMP achieves similar performance to that obtained with exhaustive search.
Overall these strategies achieve around 5 dB improvement with respect to the adaptive CS approach.
The worst performance is obtained with LS, with a big gap with respect to the other methods.
When the unquantized cluster channel model is considered, the results are different, as it can bee seen in Fig. \ref{fig:NMSE_off}.
While LS provides the same NMSE, the performance of the other approaches rapidly degrades.
Besides, the NSME of the adaptive CS and exhaustive search approach do not scale well with the SNR.
The error comes from two facts: i) the AoA/AoD of the dominant paths do not lie exactly on the grid associated with the beamformers in the codebook, ii)  the rank of the channel is higher than the number of discovered dominant paths $K$.
The way to improve the estimation of the adaptive CS is to increase the number of paths the algorithm is looking for, $K$, which implies an increase in the number of training steps $ LK^2\left(\frac{LK}{L_\text{r}}\right)\log_L \left( \frac{G_\text{r}}{K} \right)$.
Selecting an optimal $K$ for the adaptive CS and exhaustive search is another open problem.
The performance of OMP also degrades with respect to the simple channel model. 
However, the automatic stop criterion based on the residual error, allows the algorithm to look for a higher number of nonzero elements in the virtual channel model while keeping the solution sparse.
That results in a reasonable NMSE over the whole SNR range.

\subsection{Spectral efficiency with unconstrained precoding/combining and estimated channels}

\begin{figure*}[!th]
\centering
\subfloat[Simple channel model $N_\text{cl}=4$, $N_\text{ray}=1$]{\includegraphics[width=0.5\linewidth]{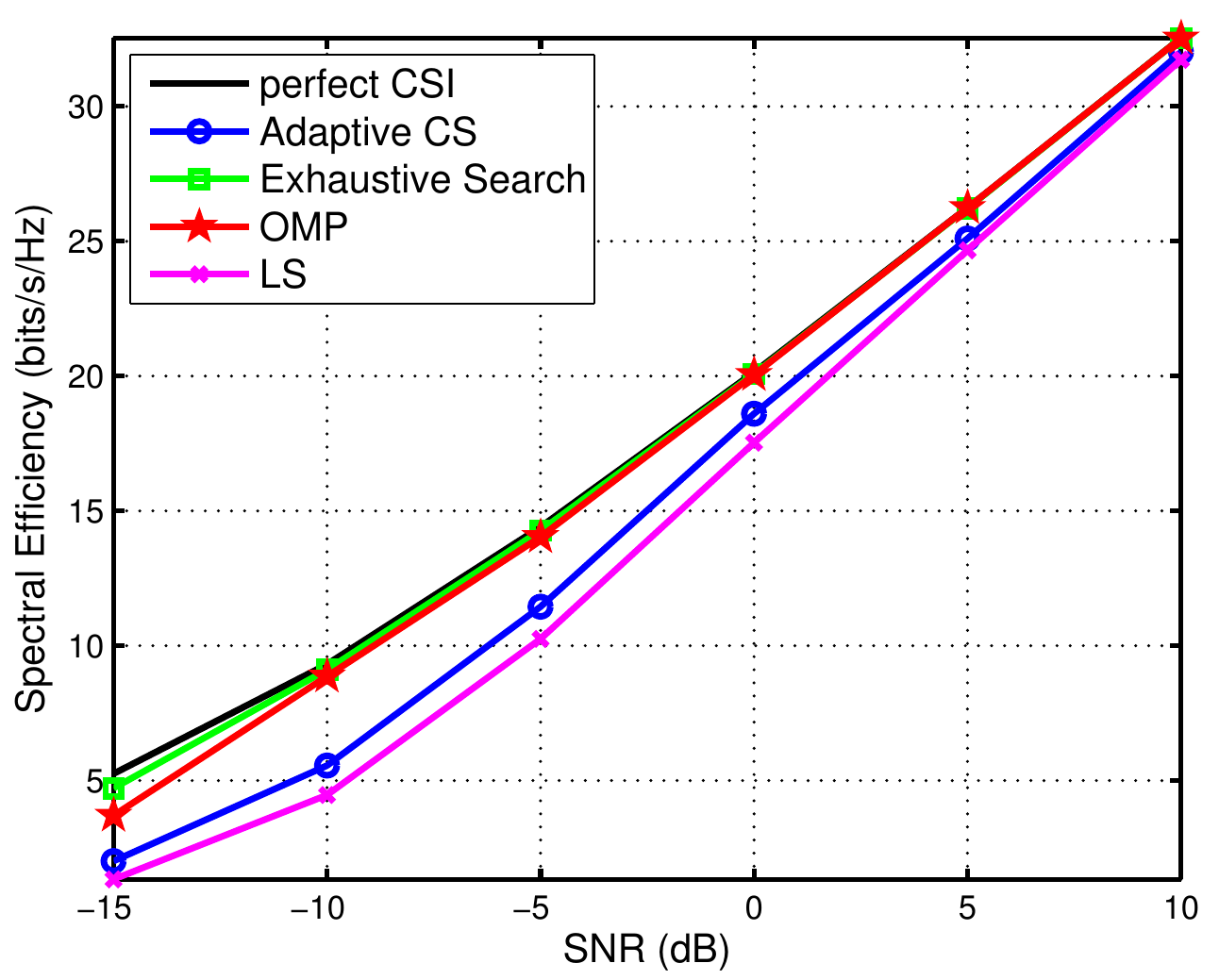}%
\label{fig:SE_on}}
\subfloat[Cluster Channel model $N_\text{cl}=4$, $N_\text{ray}=6$ ]{\includegraphics[width=0.5\linewidth]{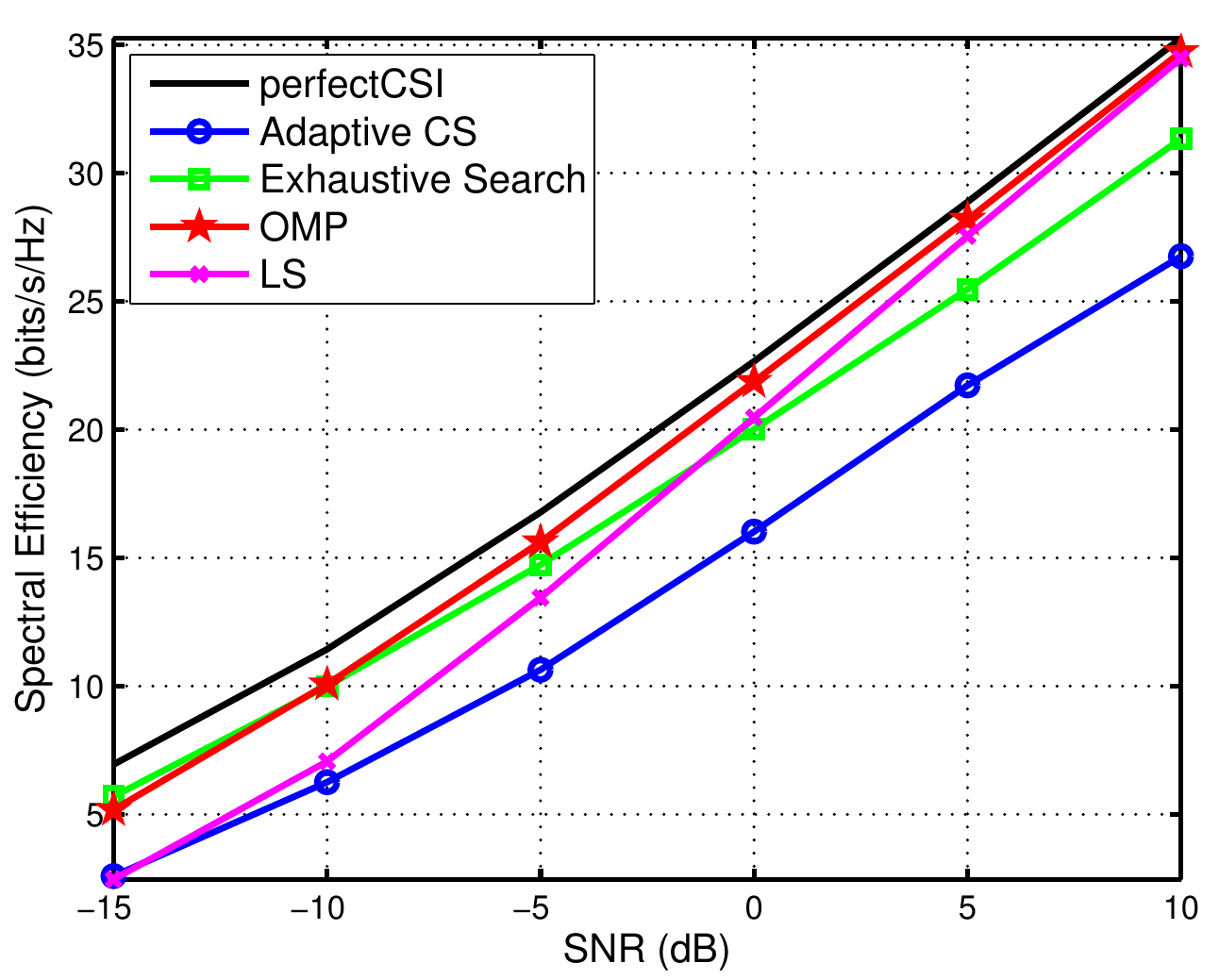}%
\label{fig:SE_off}}
\caption{Achievable spectral efficiency with the estimated channel vs SNR: $N_\text{t}=64$, $N_\text{r}=16$, $L_\text{t}=8$, $L_\text{r}=4$ and $G_\text{r}=G_\text{t}=64$. Number of training steps $M=256$.}
\label{fig:SEvsSNR}
\end{figure*}

The NMSE does not directly reflect the impact of the estimation error on the achievable spectral efficiency, which is the final aim.
To analyze that, we compute the theoretical spectral efficiency achievable with the estimated channel state information.
We consider equal power allocation and a fixed number of data streams $N_\text{s}=L_\text{r}=4$.
Here, we are interested in the theoretical bound without taking into account the constraints introduced by the hybrid architectures in the precoding/combining design.
Therefore, the optimal precoder and combiner are designed as the $N_\text{s}$ main singular vectors of the singular value decomposition of the estimated channel.
Fig. \ref{fig:SEvsSNR} shows the achievable spectral efficiency for each channel model.

\begin{figure*}[!th]
\centering
\subfloat[Simple channel model: $N_\text{cl}=4$, $N_\text{ray}=1$]{\includegraphics[width=0.5\linewidth]{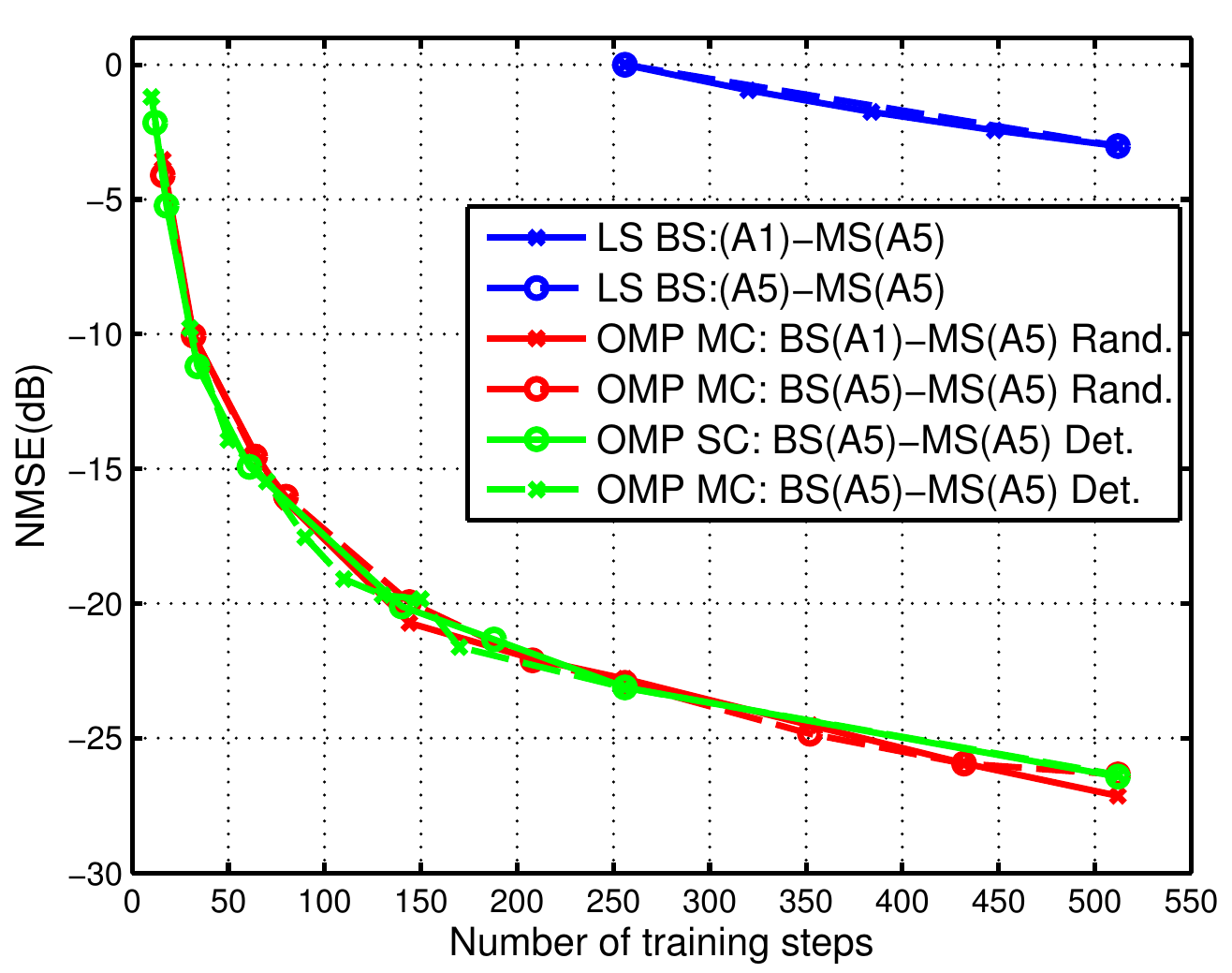}%
\label{fig:NMSEvsTS_on}}
\subfloat[Cluster channel model: $N_\text{cl}=4$, $N_\text{ray}=6$]{\includegraphics[width=0.5\linewidth]{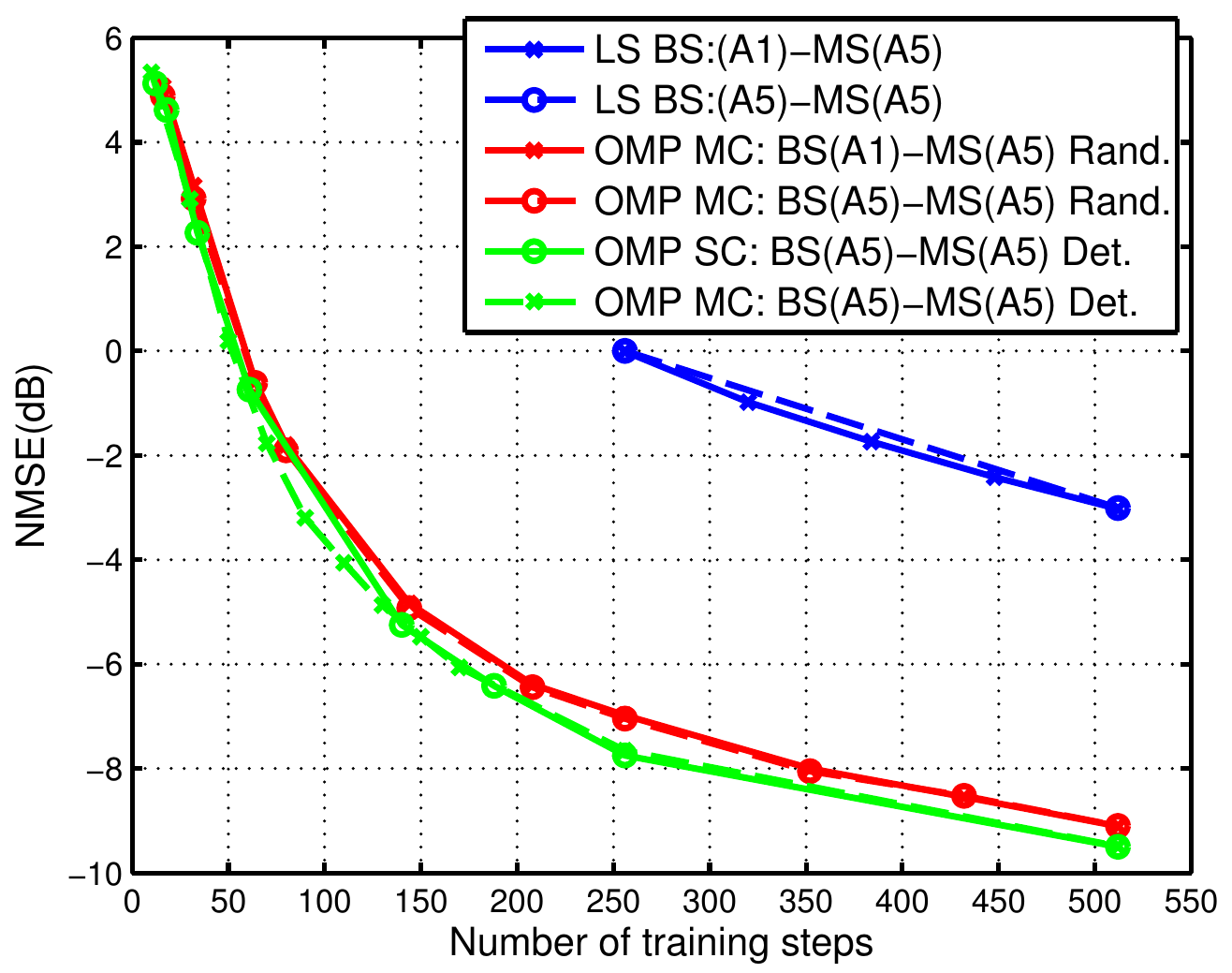}%
\label{fig:NMSEvsTS_off}}
\caption{NMSE vs number of training steps: SNR $ = 0$ dB, $N_\text{t}=64$, $N_\text{r}=16$, $L_\text{r}=4$, $G_\text{t}=64$ and $G_\text{r}=16$.}
\label{fig:NMSEvsTS}
\end{figure*}

\begin{figure*}[!th]
\centering
\subfloat[Simple channel model: $N_\text{cl}=4$, $N_\text{ray}=1$]{\includegraphics[width=0.5\linewidth]{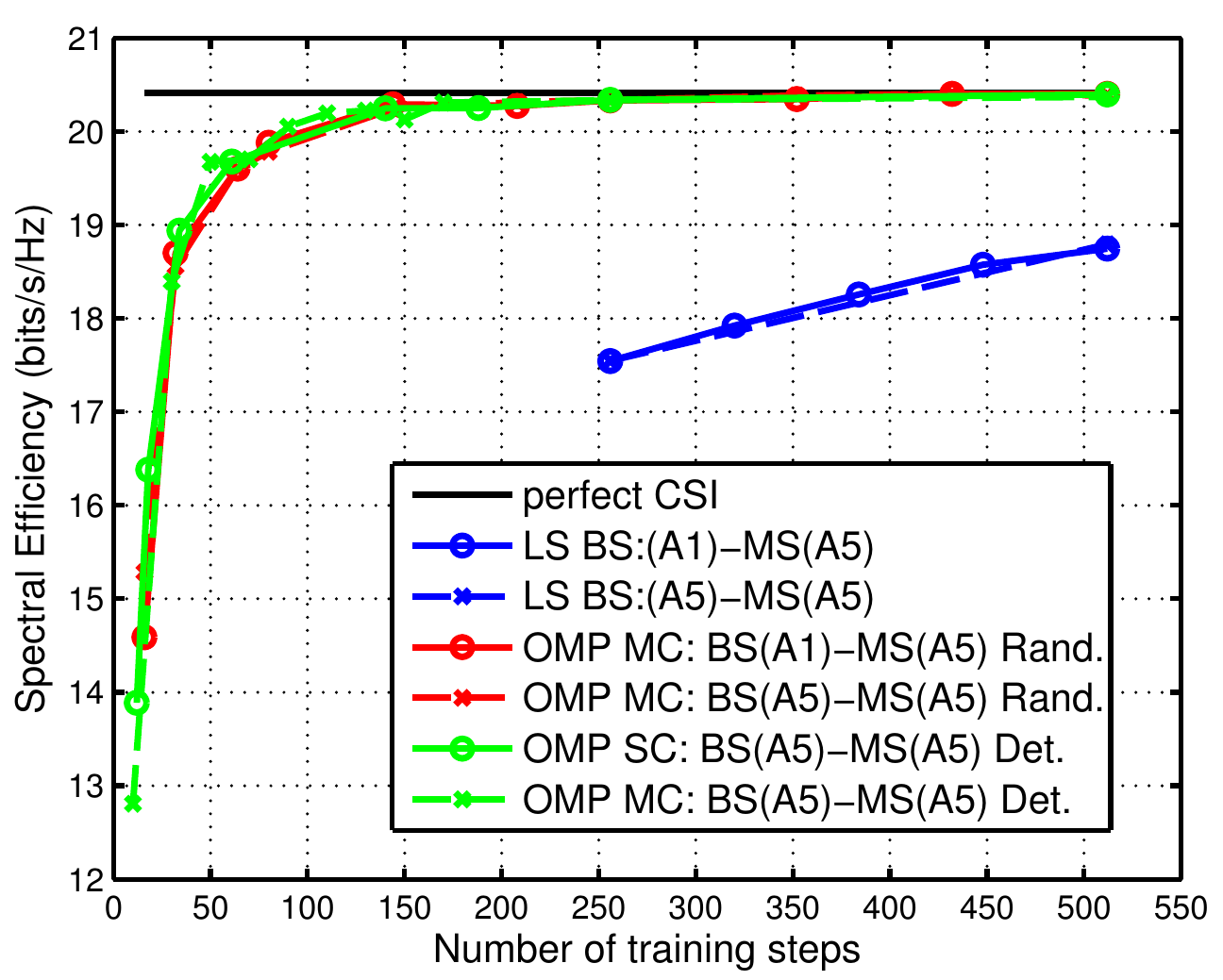}%
\label{fig:SEvsTS_on}}
\subfloat[Cluster Channel model: $N_\text{cl}=4$, $N_\text{ray}=6$]{\includegraphics[width=0.5\linewidth]{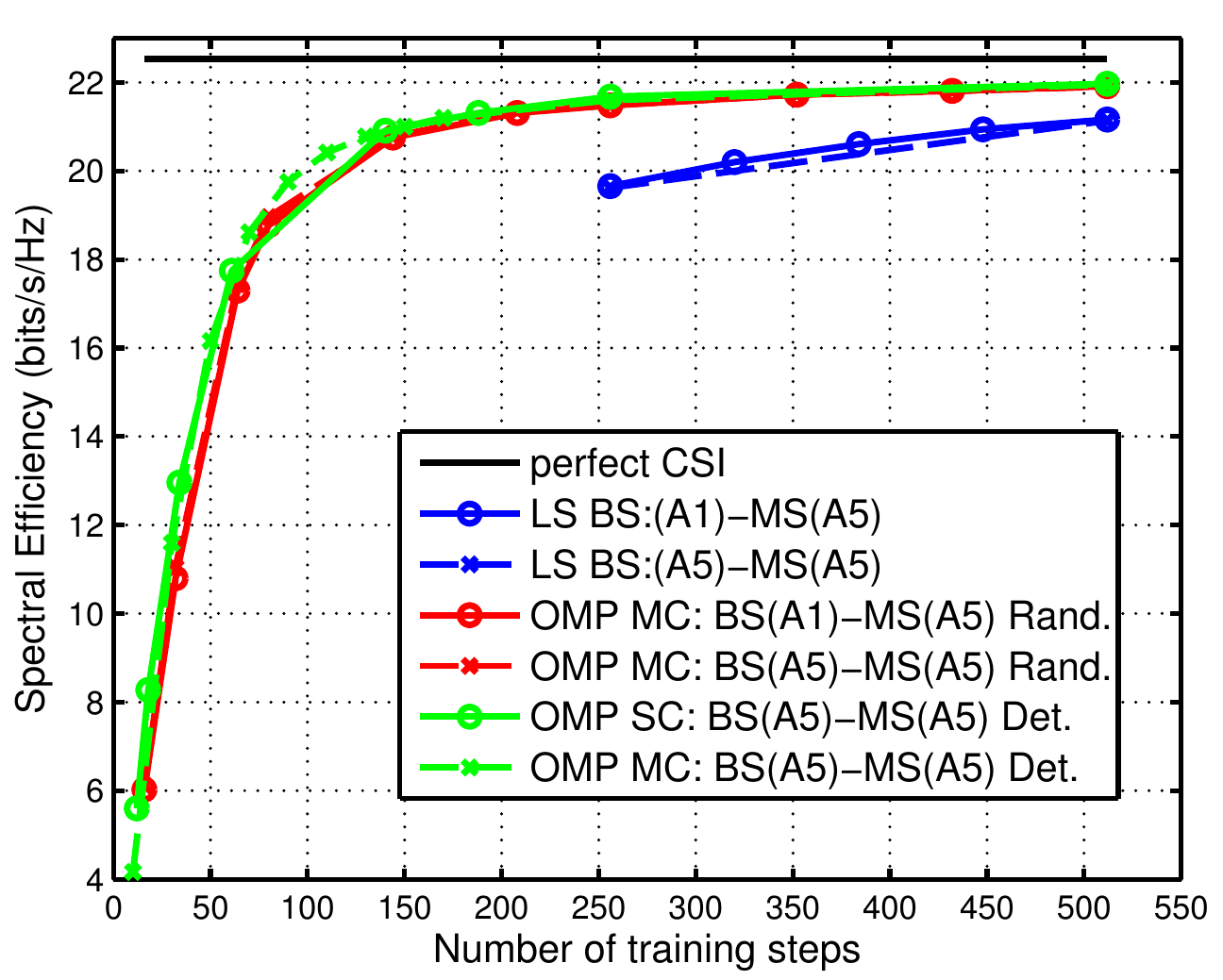}%
\label{fig:SEvsTS_off}}
\caption{Achievable spectral efficiency with the estimated channel vs number of training steps: SNR $ = 0$ dB, $N_\text{t}=64$, $N_\text{r}=16$, $L_\text{r}=4$, $G_\text{t}=64$ and $G_\text{r}=16$.}
\label{fig:SEvsTS}
\end{figure*}

The main motivation of the compressed sensing based approach is to reduce the number of training overhead needed to obtain a good estimation of the channel.
Fig. \ref{fig:NMSEvsTS} shows the NMSE obtained with OMP and LS in terms of the number of training steps.
The architectures A1 and A5 are considered for the BS and A5 for the MS.
In this case, we consider grid resolutions $G_\text{t}=64$ and $G_\text{r}=16$, i.e. the virtual channel model, to evaluate the performance of the CS based approach with the proposed deterministic training sequences in addition to the random ones.
The deterministic designs are described in section III D for the single combiner (SC) and the multiple combiner (MC) model when the architecture A5 is considered at the BS and MS.
The random sequences are generated as pseudorandom vectors with entries $\{\pm 1,\pm \mathrm{j}\}$ for A1, and binary pseudorandom vectors with exactly a one and zeros elsewhere for A5.
The training sequences for LS are designed to satisfy the orthogonality condition.
Fig. \ref{fig:NMSEvsTS} and Fig. \ref{fig:SEvsTS} show the NMSE and achievable spectral efficiencies with the proposed channel estimation strategies in terms of the training steps for SNR $=0$ dB.
The results in Fig. \ref{fig:NMSEvsTS_on} and Fig. \ref{fig:SEvsTS_on} show a big gap between OMP and LS.
With the sparse recovery approach, only $75$ training steps are required to obtain an NMSE $\approx -15$ dB and near optimal spectral efficiency.
When considering the unquantized clustered channel model, Fig. \ref{fig:NMSEvsTS_off} and Fig. \ref{fig:SEvsTS_off}, the reconstruction is worse and the gap between OMP and LS decreases.
The support of the solution can not be recovered perfectly due to the high noise power and not sufficient sparsity, and a higher number of measurements is required to obtain a good reconstruction. Aside the perfect recovery conditions based on the coherence, recent results \cite{WeiRodriguezWassell_Tight_Frames:2012} show that unit norm tight frames minimize the mean squared error (MSE) over all sparsity levels using the standard sparse recovery algorithms. This fact explains the small gap between the random and deterministic designs when sparsity is not sufficient, in the presence of noise. Although the deterministic design enjoys low mutual coherence, both designs produce tight measurement matrices that perform similarly in the MSE sense in this regime.

\subsection{Spectral efficiency with hybrid combining and estimated channels}

In this simulation, we evaluate the performance of each hybrid architecture in terms of the achievable spectral efficiency.
We consider the same downlink MIMO channel with $N_\text{t}=64$ and $N_\text{r}=16$ antennas and the cluster channel model with $N_\text{cl}=4$ clusters and $N_\text{ray}=6$ per cluster.
We assume perfect channel knowledge at the transmitter and the receiver.
The BS is implemented with a full digital beamformer architecture with one RF chain per antenna.
The unconstrained precoder is set to $\mathbf{F}=\mathbf{V}_{L_\text{r}}\mathbf{\Gamma}$, with $\mathbf{V}_{L_\text{r}}$ the main $L_\text{r}=N_\text{s}$ eigenvectors of singular value decomposition of the channel $\mathbf{H} =\mathbf{U \Sigma V}^*$, and $\mathbf{\Gamma}$ a diagonal matrix with the optimal power allocation across streams given by waterfilling.

We analyze the achievable spectral efficiency of each hybrid architecture deployed at the receiver side.
The following hybrid combiner designs are applied.
The sparse hybrid precoder algorithm \cite{Ayach2014} is used to design the hybrid combiner for A1, A2, A3, and A4 as described in Section \ref{sec:combine}.
For the hybrid antenna selection architectures, A5 and A6, we compute the achievable spectral efficiency of the system with the incremental algorithm \ref{alg:HAS}.

\begin{figure}[!t]
  \centering
  \centerline{\includegraphics[width=0.5\linewidth]{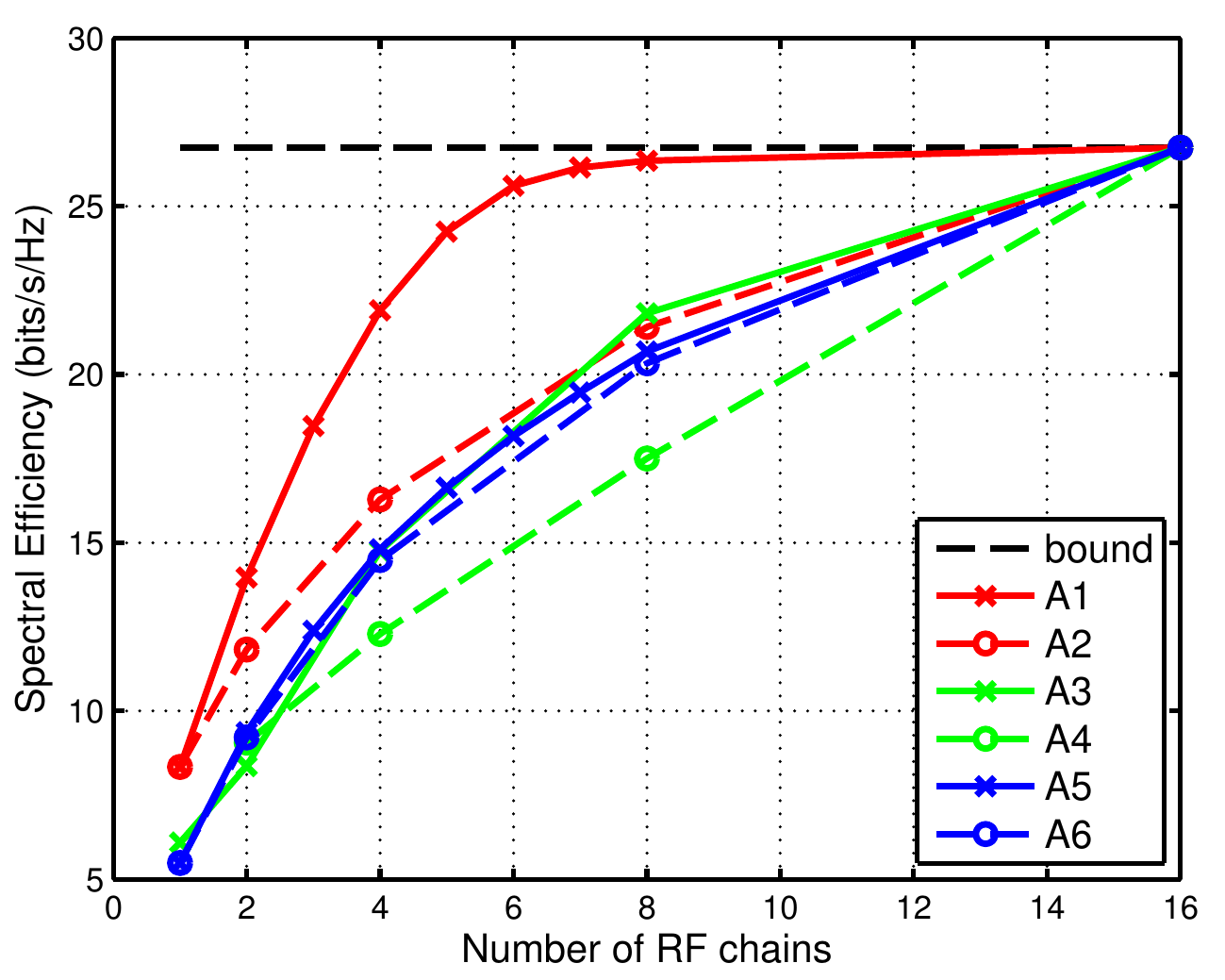}}  
  \caption{Spectral efficiency vs number of RF chains: $N_\text{t}=64$, $N_\text{r}=16$, SNR $= 0$ dB, $N_\text{s}=L_\text{r}$.}
\label{fig:SEvsRF}
\end{figure}

Fig. \ref{fig:SEvsRF} shows the spectral efficiency achieved with each architecture in terms of the number of RF chains with $N_\text{s}=L_\text{r}$ data streams.
The signal to noise ratio is fixed to SNR $=0$ dB which is expected in mmWave systems due to the large bandwidth.
As expected, the hybrid architecture A1 achieves the highest spectral efficiency with a significant gap with respect to the others.
Around $L_\text{r}=6$ RF chains are sufficient to achieve the capacity of the channel.
Despite A6 has lower hardware complexity than A5, their performance almost overlaps, both achieving similar spectral efficiencies to those obtained with A2.
The architectures based on antenna selection and combining, A3 and A4, do not result in a significant improvement with respect to simple hybrid antenna selection, A5 and A6.
While A3 slightly outperforms A2, A5 and A6 for high number of RF chains, A4 provides the worst results.
The reason of this bad performance is the specific combining algorithm used in these simulations.
In fact, A4 is a more general architecture than A6, and it should achieve at least the same spectral efficiency. Designing near optimal precoding/combining algorithms for A2, A3 and A4 is still an open problem.

\begin{figure}[!t]
  \centering
  \centerline{\includegraphics[width=0.5\linewidth]{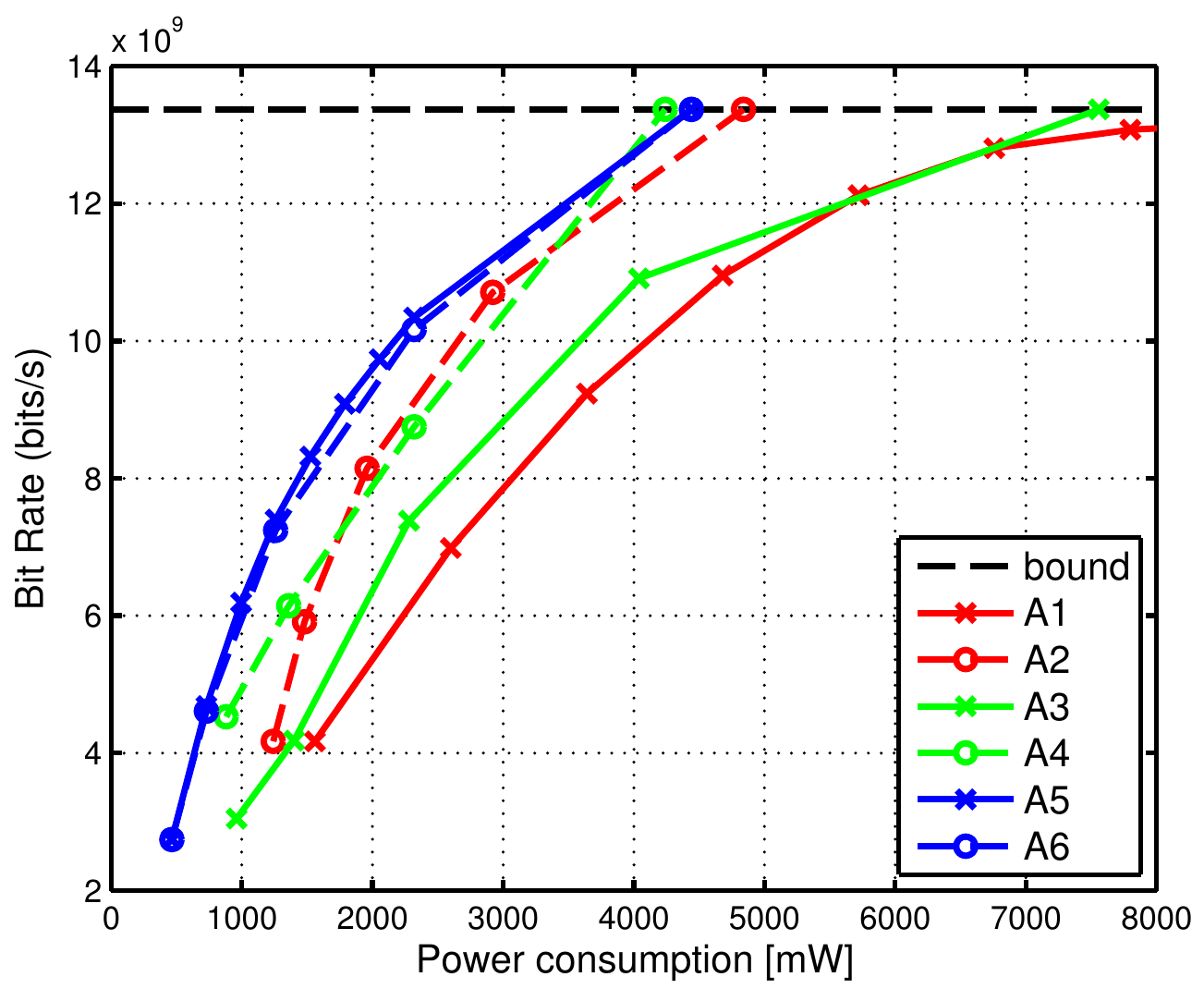}}  
  \caption{Bit rate vs power consumption: BW $= 500$ MHz, $N_\text{t}=64$, $N_\text{r}=16$, SNR $=0$ dB.}
\label{fig:BRvsP}
\end{figure}

As we have seen, there is a big difference in terms of power consumption between receivers (see Fig. \ref{powerReduction}).
Since power consumption is one of the main concerns in mmWave, it is interesting to analyze the spectral efficiencies for equal power consumption.
We represent in Fig. \ref{fig:BRvsP} the achievable bit rate in terms of the power consumption of the RF circuit.
The power consumptions are computed based on the number of RF chains and the power model in section \ref{sec:powermodel}.
The bit rate is determined from the computed spectral efficiencies and considering a bandwidth of BW $=500$ MHz.
For equal power consumption, hybrid antenna selection architectures A5 and A6 provide the highest bit rates, with a small gap with respect to architectures considering subsets of antennas, A2 and A4. 
Alternatively, the power consumption of A1 and A3 increases very rapidly with the number of RF chains, resulting in the most energy inefficient architectures. 

\section{Conclusions}\label{sec:conclusion}
In this paper we  analyzed the potential of six hybrid combining architectures for mmWave communications using phase shifters or switches as main building blocks in the analog processing stage. To complete our assessment, we developed new algorithms for channel estimation that work with all the architectures, and specific hybrid combining algorithms for the switch architecture. We investigated the trade-off of each architecture in terms of channel estimation performance, achievable spectral efficiencies, and power consumption.


The architectures based on phase shifters exploits the large antenna array gain achieving near optimal spectral efficiencies when the number of RF chains is comparable to the number of data streams.
The receivers based on switches are low power and low complexity solutions, achieving a high reduction in the power consumption per RF chain.
For equal power consumption, a higher number of RF chains can be used with switches, which results in an increased spectral efficiencies.
Numerical results show that for equal power consumption, all the architectures provide close spectral efficiencies.
Table~\ref{tab:comparison} summarizes the observations for the different architectures when using the same number of RF chains, 
emphasizing the hardware complexity- and power consumption-spectral efficiency trade-offs. 
Simulations results show that similar channel reconstruction is achieved with all the architectures and comparable to that obtained by exhaustive search.
A high number of RF chains in the mobile station reduce the number of needed training steps at expenses of increasing the power consumption of the circuit.

\begin{table}[htbp]  
  \centering  
    \begin{tabular}{|l|c|c|c|c|}
		\hline 
    Architecture   & Power  & Hardware  & NMSE in & Spectral  \\
     & consumption & complexity & channel estimation & efficiency \\
		\hline
		A1 & High & High & Low & High \\ \hline
		A2 & Medium & Medium & Low & Medium \\ \hline
		A3 & Medium & High & Low & Medium \\ \hline
		A4 & Low & Medium & Low & Medium \\ \hline
		A5 & Low & Low & Low & Medium \\ \hline
		A6 & Low & Low &  Low & Medium \\
		\hline				
    \end{tabular}%
	\caption{Power consumption and hardware complexity versus performance of the different architectures for a fixed number of RF chains.}			
	\label{tab:comparison}%
\end{table}%

It is of interest to extend these results using a more complete system model, including impairments of mmWave devices like noise figure or insertion losses, which  impact the actual  SNR at the input of the digital combiner. A switches-based hybrid architecture can also be considered for the transmitter, with the resulting joint selection of the subset of antennas at transmitter and receiver.


\end{document}